%% file: main.tex
\documentclass[aps,prd,twocolumn,superscriptaddress,showpacs,preprintnumbers,amsmath,amssymb,floatfix,nofootinbib]{revtex4-2}

\usepackage[dvipdfmx]{graphicx}
\usepackage{graphicx}
\usepackage{dcolumn}
\usepackage{bm}
\usepackage{multirow}
\usepackage{color}
\usepackage{url}
\usepackage{xcolor}
\usepackage[italic]{hepnames}
\usepackage{tabularx}
\usepackage{array,booktabs}
\usepackage{caption}
\usepackage{subcaption}
\usepackage{adjustbox}
\usepackage{placeins}
\usepackage{lineno}
\usepackage{orcidlink}

\def\B       {\ensuremath{B}\xspace}
\def\Bbar    {\kern 0.18em\overline{\kern -0.18em B}{}\xspace}
\def\Bb      {\ensuremath{\Bbar}\xspace}
\def\BB      {\ensuremath{B\Bbar}\xspace} 
\def\Bz      {\ensuremath{B^0}\xspace}
\def\Bzb     {\ensuremath{\Bbar^0}\xspace}
\def\BzBzb   {\ensuremath{\Bz {\kern -0.16em \Bzb}}\xspace}
\def\Bu      {\ensuremath{B^+}\xspace}
\def\Bub     {\ensuremath{B^-}\xspace}

\def\BpBm    {\ensuremath{\Bu {\kern -0.16em \Bub}}\xspace}

\def\nub        {\ensuremath{\overline{\nu}}\xspace}

\def\qbar  {\ensuremath{\overline q}\xspace}

\definecolor{BlueViolet}{rgb}{0.2, 0.00, 0.7}
\definecolor{Blue}{rgb}{0.15, 0.00, 0.9}

\newcommand{\brkaonkpi}{(2.154\pm0.089\pm0.078)\times 10^{-4}}
\newcommand{\brkaonkthreepi}{(2.287\pm0.088\pm0.093)\times 10^{-4}}
\newcommand{\brkaoncomb}{(2.221\pm0.063\pm0.077)\times 10^{-4}}

\newcommand{\brkaoncombabs}{(2.22\pm0.06\pm0.08)\times 10^{-4}}

\newcommand{\brpionkpi}{(2.607\pm0.023\pm0.083)\times 10^{-3}}
\newcommand{\brpionkthreepi}{(2.640\pm0.022\pm0.101)\times 10^{-3}}
\newcommand{\brpioncomb}{(2.623\pm0.016\pm0.086)\times 10^{-3}}

\newcommand{\brpioncombabs}{(2.62\pm0.02\pm0.09)\times 10^{-3}}

\newcommand{\ratiokpi}{(8.26\pm0.35\pm0.16)\times 10^{-2}}
\newcommand{\ratiokthreepi}{(8.56\pm0.34\pm0.16)\times 10^{-2}}
\newcommand{\ratiocomb}{(8.41\pm0.24\pm0.13)\times 10^{-2}}

\newcommand{\ratiocombabs}{(8.41\pm0.24\pm0.13)\times 10^{-2}}

\newcommand{\aOneRatio}{1.066\pm0.042\pm0.018\pm0.023}

\newcommand{\aOnePion}{0.884\pm0.004\pm0.003\pm0.016}
\newcommand{\aOneKaon}{0.913\pm0.019\pm0.008\pm0.013}

\newcommand{\pionmodemath}{\Bb^0 \to D^{*+} \pi^-}
\newcommand{\kaonmodemath}{\Bb^0 \to D^{*+} K^-}

\newcommand{\pionmode}{\ensuremath{\Bb^0 \to D^{*+} \pi^-}\xspace}
\newcommand{\kaonmode}{\ensuremath{\Bb^0 \to D^{*+} K^-}\xspace}

\newcommand{\ktwopimath}{D^{0} \to K^{-} \pi^{+}}
\newcommand{\kthreepimath}{D^{0} \to K^{-} 2\pi^{+} \pi^{-}}

\newcommand{\tev}{\ensuremath{\mathrm{\,Te\kern -0.1em V}}\xspace}
\newcommand{\gev}{\ensuremath{\mathrm{\,Ge\kern -0.1em V}}\xspace}
\newcommand{\mev}{\ensuremath{\mathrm{\,Me\kern -0.1em V}}\xspace}
\newcommand{\kev}{\ensuremath{\mathrm{\,ke\kern -0.1em V}}\xspace}
\newcommand{\ev}{\ensuremath{\mathrm{\,e\kern -0.1em V}}\xspace}
\newcommand{\gevc}{\ensuremath{{\mathrm{\,Ge\kern -0.1em V\!/}c}}\xspace}
\newcommand{\mevc}{\ensuremath{{\mathrm{\,Me\kern -0.1em V\!/}c}}\xspace}
\newcommand{\gevcc}{\ensuremath{{\mathrm{\,Ge\kern -0.1em V\!/}c^2}}\xspace}
\newcommand{\mevcc}{\ensuremath{{\mathrm{\,Me\kern -0.1em V\!/}c^2}}\xspace}

\begin{document}


\title{Measurements of the branching fractions $\mathcal{B}(\Bb^0 \to D^{*+} \pi^-)$ and $\mathcal{B}(\Bb^0 \to D^{*+} K^-)$ and tests of QCD factorization}

\noaffiliation
 


\input{pub611-orcid.tex}

\begin{abstract}
		Using $(771.6 \pm 10.6) \times 10^6$ \BB meson pairs recorded by the Belle experiment at the KEKB $e^+ e^-$ collider, we report the branching fractions $\mathcal{B}(\Bb^0 \to D^{*+} \pi^-) = \brpioncombabs$ and $\mathcal{B}(\Bb^0 \to D^{*+} K^-) = \brkaoncombabs$; the quoted uncertainties are statistical and systematic, respectively. 
		A measurement of the ratio of these branching fractions is also presented, $\mathcal{R}_{K/\pi}=\mathcal{B}(\Bb \to D^{*+}K^-) /\mathcal{B}( \Bb \to D^{*+}\pi^-) = \ratiocombabs$, where systematic uncertainties due to the $D^{*+}$ meson reconstruction cancel out.
		Furthermore, we report a new QCD factorization test based on the measured ratios for $\Bb\to D^{*+}h^-$ and $\Bb\to D^{*+}\ell^-\nu$ decays at squared momentum transfer values equivalent to the mass of the $h = \pi$ or $K$ hadron. 
		The parameters $|a_1(h)|$ are measured to be $|a_1(\pi)|=\aOnePion$ and $|a_1(K)|=\aOneKaon$, where the last uncertainties account for all external inputs. These values are approximately 15\% lower than those expected from theoretical predictions.
		Subsequently, flavor $SU(3)$ symmetry is tested by measuring the ratios for pions and kaons, $|a_1(K)|^2/ |a_1(\pi)|^2=\aOneRatio$, as well as for different particle species. The ratio is consistent with unity and therefore no evidence for $SU(3)$ symmetry breaking effects is found at the $5\%$ precision level. 
\end{abstract}

{
\let\clearpage\relax
\maketitle
}

\section{Introduction}\label{chapter:measurement}

Hadronic \B decays such as $\Bb^{0} \to D^{*+} h^{-}$, where $h$ denotes a pion or kaon, are interesting for a variety of reasons. Their branching fractions are large and therefore large data samples containing them are available for precision measurements.
Since the $\Bb^{0} \to D^{*+} K^{-}$ decay involves virtual $b \to cW^-$ and $W^-\to \bar{u} s$ transitions, its branching fraction is approximately five times smaller than that of $\Bb^{0} \to D^{*+} \pi^{-}$, which proceeds via a $W^-  \to \bar{u} d$ transition.
The branching fractions of both decays allow for precision tests of the theoretical framework used to calculate hadronic \B decays as well as to constrain physics beyond the Standard Model (SM)~\cite{Bordone:2020gao,Lenz:2019lvd}. The branching fractions of these processes were measured by many experiments such as CLEO~\cite{Bortoletto:1991kz, Bebek:1987bp, Giles:1984yg, Alam:1994bi,CLEO:1997vmd}, OPAL~\cite{Akers:1994qg}, ARGUS~\cite{ARGUS:1986ckc, Albrecht:1986hs, Albrecht:1990cs}, and more recently by BaBar~\cite{Aubert:2006jc, BaBar:2006rof, Aubert:2005yt} and Belle~\cite{Abe:2001waa}.  
The ratios of branching fractions allow for a probe into the symmetries of the SM, such as flavor $SU(3)$, with cancellation of the major systematic uncertainties, e.g.\ those from $D^{*+}$ reconstruction. Recent measurements of these ratios were reported by BaBar~\cite{Aubert:2005yt}, Belle~\cite{Abe:2001waa}, and LHCb~\cite{Aaij:2013xca}.

Using the semileptonic decay rate $d\Gamma(\Bb^0\to D^{*+} \ell \nub)/dq^{2}$ at a fixed lepton-momentum transfer, $q^{2}=m^2_{h}$, combined with the $\Bb^{0} \to D^{*+} h^{-}$ decay rate, one can measure $|a_1(q^{2})| \equiv |a_1(h)|$, a fundamental parameter in the description of hadronic \B decays \cite{Beneke:2000_qcd_factorisation}. One finds
\begin{equation}\label{eq:diff_decay_rate}\begin{split}
\Gamma(\Bb^0 \to D^{*+}  h^{-}) = & 6\pi^{2}  \tau_B |V_{uq}|^{2} f_{h}^{2}  X_{h} |a_1(q^2)|^{2} \times \\
& d\Gamma(\Bb^0 \to D^{*+}  \ell^- \nub ) /dq^2 |_{q^{2}=m^2_{h}},
\end{split}
\end{equation}
where $\tau_B$ is the lifetime of the $B^{0}$ meson, $V_{uq}$ the CKM matrix element, $f_{h}$ the decay constant of the respective meson, and $X_{h}=1 + \mathcal{O}(m_{h}^{2}/m_{B}^{2})$.
A measurement of $|a_1(q^{2})|$ requires determinations of hadronic and semileptonic branching fractions and has never been performed by a single experiment. Measurements based on results from different experimental sources are in tension with the theoretical predictions~\cite{Fleischer:2010ca}. The semileptonic inputs for our measurement are taken from Refs.~\cite{Waheed_2019, Ferlewicz:2020lxm}. Charge conjugation is implied throughout this paper.

\subsection{The Belle detector and data sample}

The results use the full $\Upsilon(4S)$ data sample containing $(771.6 \pm 10.6) \times 10^6$ \BB meson pairs recorded with the Belle detector~\cite{Abashian:2000cg,Belle:2012iwr} at the KEKB asymmetric-energy $e^+e^-$ collider~\cite{Kurokawa:2001nw,Abe:2013kxa}. The subdetectors relevant for our study are: a silicon vertex detector, a 50-layer central drift chamber (CDC), an array of aerogel threshold Cherenkov counters (ACC), a barrel-like arrangement of time-of-flight scintillation counters (TOF), and an electromagnetic calorimeter comprised of CsI(Tl) crystals. All these components are located inside a superconducting solenoid coil that provides a 1.5 T magnetic field. The $z$-axis is the direction opposite to the $e^+$ beam.

Monte Carlo (MC) simulation studies are performed using a sample corresponding to five times the integrated luminosity of this dataset. The MC sample is generated using the EVTGEN~\cite{Ryd:2005zz}, PYTHIA~\cite{Sj_strand_2006}, and PHOTOS~\cite{Barberio:1993qi} packages with interference effects due to final-state radiation switched on. We reconstruct candidate events using the Belle II analysis software framework~\cite{Kuhr_2018}, after converting them to the Belle II data format with the B2BII package~\cite{Gelb_2018}.

\section{Measurement of $\mathcal{B}(\Bb^{0} \to D^{*+} \pi^{-})$ and $\mathcal{B}(\Bb^{0} \to D^{*+} K^{-})$ }

\subsection{Strategy and event reconstruction}\label{sec:precuts}
We reconstruct $\Bb^{0} \to D^{*+} h^{-}$ decays with a pion mass hypothesis for the charged hadron accompanying the $D^{*+}$, which we will refer to as the bachelor hadron. The sample is split into \pionmode enhanced and \kaonmode enhanced subsamples by suitably requiring kaon-pion identification criteria for the bachelor hadron. 
As the \kaonmode decay is reconstructed with the pion mass hypothesis it peaks approximately $48$ \mev lower in the energy-difference variable, $\Delta E = E^{*}_{B} - E^{*}_{\rm {beam}}$, where $E^{*}_{B}$ is the energy of the \B meson and  $E^{*}_{\rm {beam}}$ is the beam energy, evaluated in the center-of-mass frame (denoted by the symbol $^*$). Thus, peaks from both decays can be fit simultaneously, which allows the distribution in a given enhanced subsample to constrain the shape of the distribution of the corresponding depleted subsample where it is treated as a background. 

We consider $D^{*+}$ candidates from $D^{*+}\to D^0 \pi^+$ decays reconstructed from two specific $D^0$ decay channels: the highly pure but smaller branching fraction $D^0\to K^- \pi^+$ channel and the less pure but larger branching fraction $D^0\to K^- 2\pi^+ \pi^-$ channel. When accounting for the efficiencies, the expected yields of the two $D^0$ channels are of the same order.
This is a blind analysis in which the measurement is first optimized using MC simulation and then performed on data with the same selection criteria. Efficiency differences between data and MC simulation for the reconstruction of low-momentum `slow' pions from $D^{*+}\to D^0 \pi^+$ decays and particle identification are corrected for using control sample measurements.

Charged particle tracks originating from $e^+e^-$ collisions are selected by requiring the track impact parameter along the $z$ axis to be $|dz|< 4$ cm and a radial distance to the interaction point of $|dr| < 2$ cm. 
Information from the CDC, ACC and TOF is used to determine a $K/\pi$ likelihood ratio ${\mathcal{L}}_{K/\pi}={\mathcal{L}}_{K}/({\mathcal{L}}_{\pi}+{\mathcal{L}}_{K})$ for charged particle identification, where ${\mathcal{L}}_{K}$ and ${\mathcal{L}}_{\pi}$ are the likelihoods that a particular track is either a kaon or a pion, respectively.
For all high-momentum pions ($p_{\rm T} > 200 \mevc$), we require ${\mathcal{L}}_{K/\pi} < 0.6$, which is referred to as $\pi$-ID. Slow pions ($p_{\rm T} \le 200 \mevc$) from $D^{*+}\to D^0 \pi^+$ decays are excluded from these requirements since they have only limited particle identification information. For all kaons, the opposite requirement of ${\mathcal{L}}_{K/\pi} \ge 0.6$ is applied, also referred to as $K$-ID. The $D^0$ meson candidates are required to have an invariant mass, $M_{D^0}$, within the range $\mu_{D^0} - 3\sigma_{D^0} < M_{D^0} < \mu_{D^0} + 3\sigma_{D^0}$. The central value, $\mu_{D^0}$, is found by a fit to data, where the width, $\sigma_{D^0}$, is defined as the weighted average of the widths of a double Gaussian function used for the signal probability distribution function (PDF). The values of $\sigma_{D^0}$ are approximately $6$ and $7$ \mevcc in the $D^0\to K^- \pi^+$ and $D^0\to K^- 2\pi^+ \pi^-$ channels, repsectively. For the reconstruction of $D^{*+}$~candidates we use an asymmetric window for the variable $\Delta M_{D^{*+}} = M_{D^{*+}} - M_{D^0}$ of $ \mu_{\Delta M} - 3\sigma_{{\rm{left}}} < \Delta M_{D^{*+}} < \mu_{\Delta M} + 3\sigma_{{\rm right}}$, where the widths $\sigma_{{\rm{left}}}, \sigma_{{\rm{right}}}$ are based on weighted averages of a Gaussian and an asymmetric Gaussian function. The widths are approximately 0.7 \mevcc.
For $B^0$ meson candidates we require the beam-energy constrained mass to be $M_{{\rm bc}}= \sqrt{E^{*2}_{\rm{beam}}/c^4 - p^{*2}_{B}/c^2 } > 5.27 \gevcc$, where $p^{*}_{B}$ is the momentum of the \B meson in the center-of-mass frame, and the energy difference to be $-150 \mev < \Delta E < 125 \mev$. The latter is a relatively wide window chosen to simultaneously select both \pionmode and \kaonmode decays.

After applying the above selection criteria, multiple $D^{*+}$ candidates are found in approximately 2\% of candidate \B events. To select the best $D^{*+}$ candidate, a minimal $\chi^{2}$ based approach is used, with the $\chi^{2}$ defined as
\begin{equation}
\begin{split}
    \chi^2 &= \left(\frac{ M_{D^0} - m_{D^0}}{\delta_{D^0}}\right)^2 \\
    &+ \left(\frac{ \Delta M_{D^{*+}} - \Delta m_{D^{*+}} }{\delta_{\Delta M}} \right)^2.
\end{split}
\end{equation}
Here, $m_{D^0}$ denotes the world-average mass of the $D^0$ meson and $\Delta m_{D^{*+}}$ is the difference between the world-average $D^{*+}$ and $D^0$ masses~\cite{PDG}. The terms $\delta_{D^0}$ and $\delta_{\Delta M}$ are the uncertainty in $M_{D^0}$ and $\Delta M_{D^{*+}}$, respectively, propagated from the uncertainty in the vertex position, momentum and energy of the decay products within the Belle detector. If two candidates have the same $\chi^2$ value, one is chosen arbitrarily.

To correct for data-MC differences in the kaon-pion separation, control samples of $D^{*+}\to D^0(\to K^-\pi^+)\pi^+$ and $K^{0}_{\rm{S}}\to \pi^+\pi^-$ decays are used. In the $D^{*+}$ sample, efficiencies are obtained by fitting the $\Delta M$ distributions with and without particle identification criteria using loose track selection, requiring that they originate from near the interaction point. For the $K^{0}_{\rm{S}}$ sample, a loose track selection is required followed by a requirement that the momentum vector of the $K^{0}_{\rm{S}}$ and the vector pointing from the interaction point to the decay vertex align. The efficiencies are determined in a simultaneous fit to the $K^{0}_{\rm{S}}$ invariant mass distributions for candidates that pass and fail the particle identification selection. The efficiencies are calculated in bins of track polar angle and momentum. In the regions covered by the $D^{*+}$ sample, these results are used, otherwise the results from the $K^{0}_{\rm{S}}$ sample are taken. If no corrections are available for a given polar angle and momentum then the event is not included in the analysis.
Data-MC differences in the slow pion efficiencies are also corrected, and described in detail elsewhere~\cite{Waheed_2019}.
The final reconstruction efficiencies include corrected particle identification efficiencies and are found to be $\epsilon(\pionmodemath)=(32.67\pm 0.12)\% $ and $\epsilon(\kaonmodemath)=(28.33\pm0.42)\% $ for the $\ktwopimath$ channel, and $\epsilon(\pionmodemath)=(17.85\pm 0.06)\% $ and $\epsilon(\kaonmodemath)=(14.98\pm0.20)\% $ for the $\kthreepimath$ channel.

\subsection{Background}\label{background}
The remaining sources of background are from other \B meson decays and from continuum quark-pair production processes ($e^+e^-\to q\qbar$), where $q$ denotes a light-flavor or (predominantly) charm quark. 

For the $\ktwopimath$ channel, in the \pionmode sample, continuum processes account for $70\%$ of the background while the largest contributions to the background from other $B$ meson decays are from $\Bb^0 \to D^{*+} \ell^- \nub $ ($\approx 8\%$), $\Bb^0 \to D^{*+} \rho^- $ ($\approx 7\%$) and inclusive $\Bb^0 \to D^{*0} X$ ($\approx 4\%$). For the \kaonmode sample in the same $D^0$ channel, continuum processes account for $90\%$ of the background and the largest contributions to the background from other $B$ meson decays are from inclusive $\Bb^0 \to D^{*0} X$ ($\approx 2\%$) and $\Bb^0 \to D^{*+} \rho^- $ ($\approx 2\%$).

For the $\kthreepimath$ channel, in the \pionmode sample, continuum processes account for $60\%$ of the background while the largest $B$ meson decay contributions are from $\Bb^0 \to D^{*+} \rho^-$ ($\approx 13\%$), mis-reconstructed $D^0$ candidates ($\approx 10\%$) and $\Bb^0 \to D^{*+} \ell^- \nub$ ($\approx8\%$). For the \kaonmode sample in the same $D^0$ channel, continuum processes account for $85\%$ of the background and the largest $B$ meson decay contributions are from mis-reconstructed $D^0$ candidates ($\approx 5\%$) and $\Bb^0 \to D^{*+} \rho^-$ ($\approx 3\%$).

\subsection{Signal extraction}
The signal yields are determined by a simultaneous unbinned maximum-likelihood fit to the pion-enhanced and depleted samples in $\Delta E$, where the same signal PDFs are used in both samples. 
For the \pionmode decay, the signal PDF is modeled by a sum of two Gaussians and a Crystal Ball function \cite{cb_function}, while the \kaonmode decay uses the sum of a single Gaussian and a Crystal Ball function. The yields, means, and a width resolution parameter common to both modes are allowed to float. The widths of the respective channels are fixed to their MC values, but allowed to float through the common resolution factor, $\beta$, which is simultaneously fit, i.e. $\sigma_{i}^{{\rm data}} = \beta \times \sigma_{i}^{{\rm MC}} $ for each PDF $i$. The ratios of the Gaussian and Crystal Ball contributions are fixed, which introduces a small bias $(<0.5\%)$, incorporated as a source of systematic uncertainty. 

Continuum background contributions are parameterized with a second-order Chebyshev polynomial where its coefficients are fixed based on fits to MC and verified using an $M_{bc} < 5.27 \gevcc$ sideband. The yield remains free to float in the fit.

The background from \B meson decays is mostly combinatorial, with a small component peaking away from the signal region in $\Delta E$. It is thus described with a combination of PDFs for each category defined in Sec.~\ref{background}. Each component is parameterized with the sum of a Gaussian and a Crystal Ball function, and a single yield is floated for their combined PDF.  

\begin{figure*}[htb]
	\centering
	\begin{subfigure}{0.45\linewidth}
		\centering
		\includegraphics[width=\textwidth]{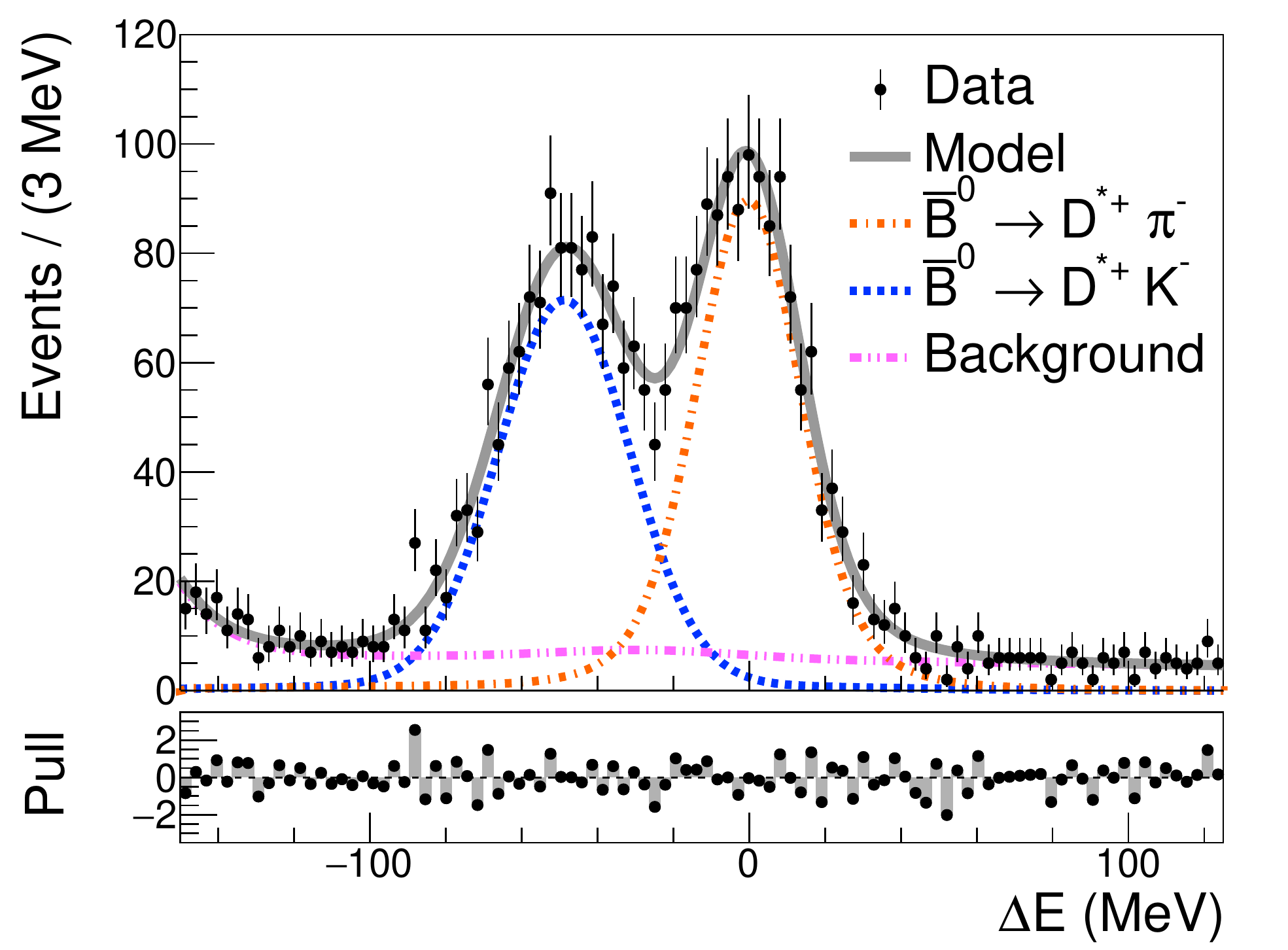}
		\caption{}
		\label{fig:final_results_fit_kpi_kaon}
	\end{subfigure}
	\begin{subfigure}{0.45\textwidth}
		\centering
        \includegraphics[width=\textwidth]{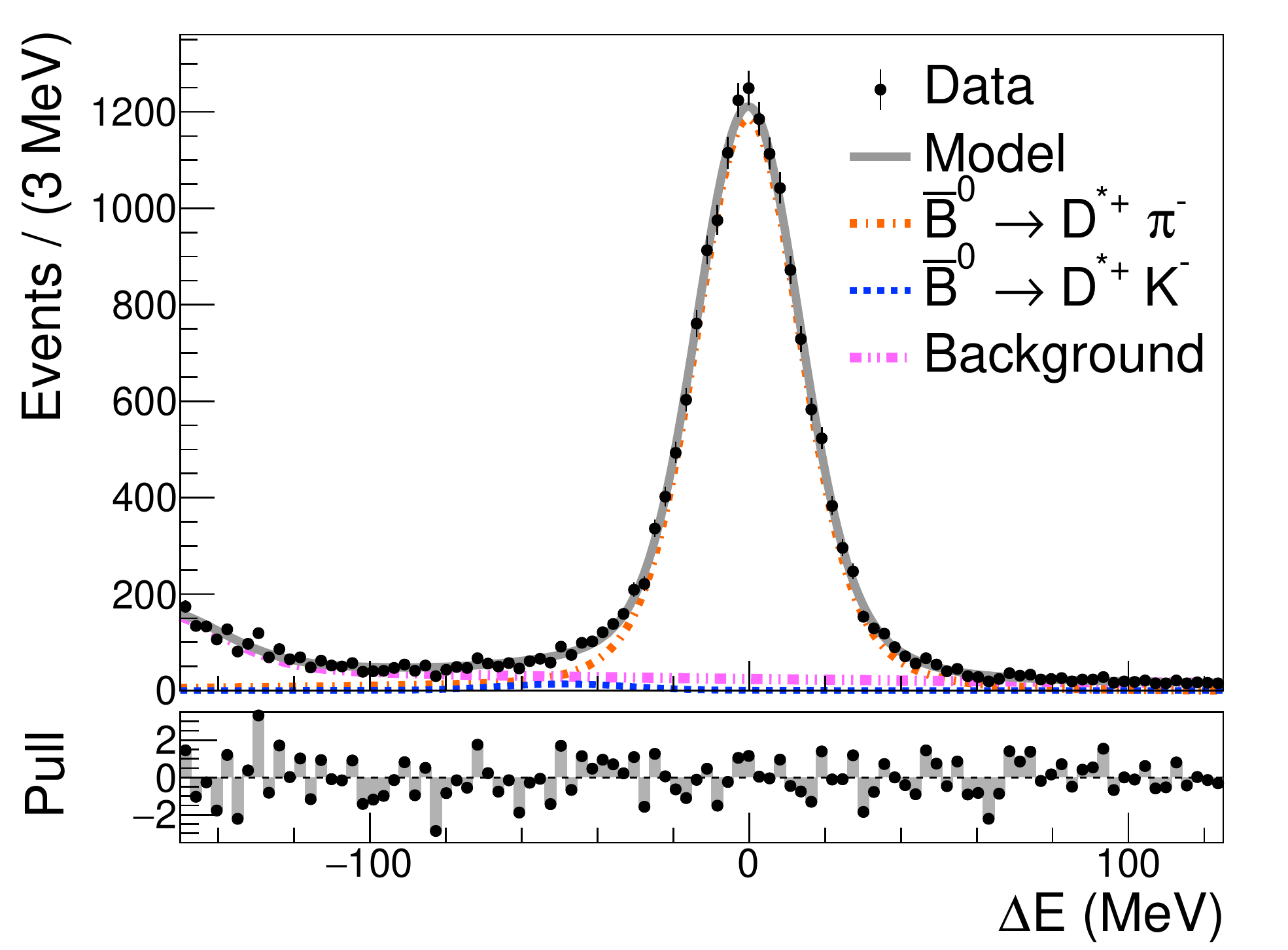}
		\caption{}
		\label{fig:final_results_fit_kpi_pion}
	\end{subfigure}
	\caption{ 
		Results of the fits to the $\Delta E$ distributions in the $D^0\to K^-\pi^+ $ channel of (a) $\Bb^0 \to D^{*+} K^-$ and (b) $\Bb^0 \to D^{*+} \pi^-$.  
	}
	\label{fig:final_fits_to_data_kpi}
\end{figure*}

\begin{figure*}[htb]
	\centering
	\begin{subfigure}{0.45\textwidth}
		\centering
		\includegraphics[width=\textwidth]{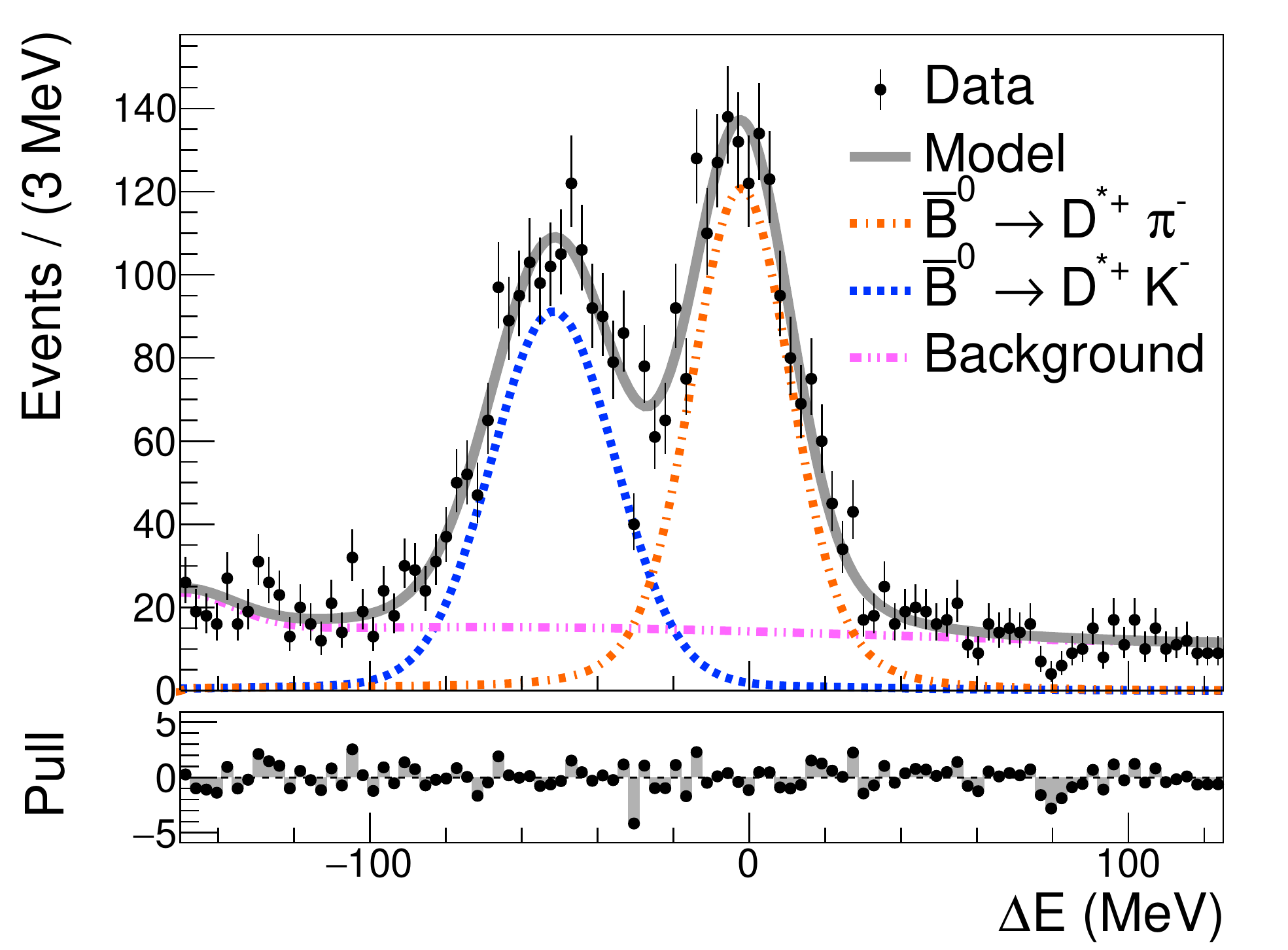}
		\caption{}
		\label{fig:final_results_fit_k3pi_kaon}
	\end{subfigure}
	\begin{subfigure}{0.45\textwidth}
		\centering
		\includegraphics[width=\textwidth]{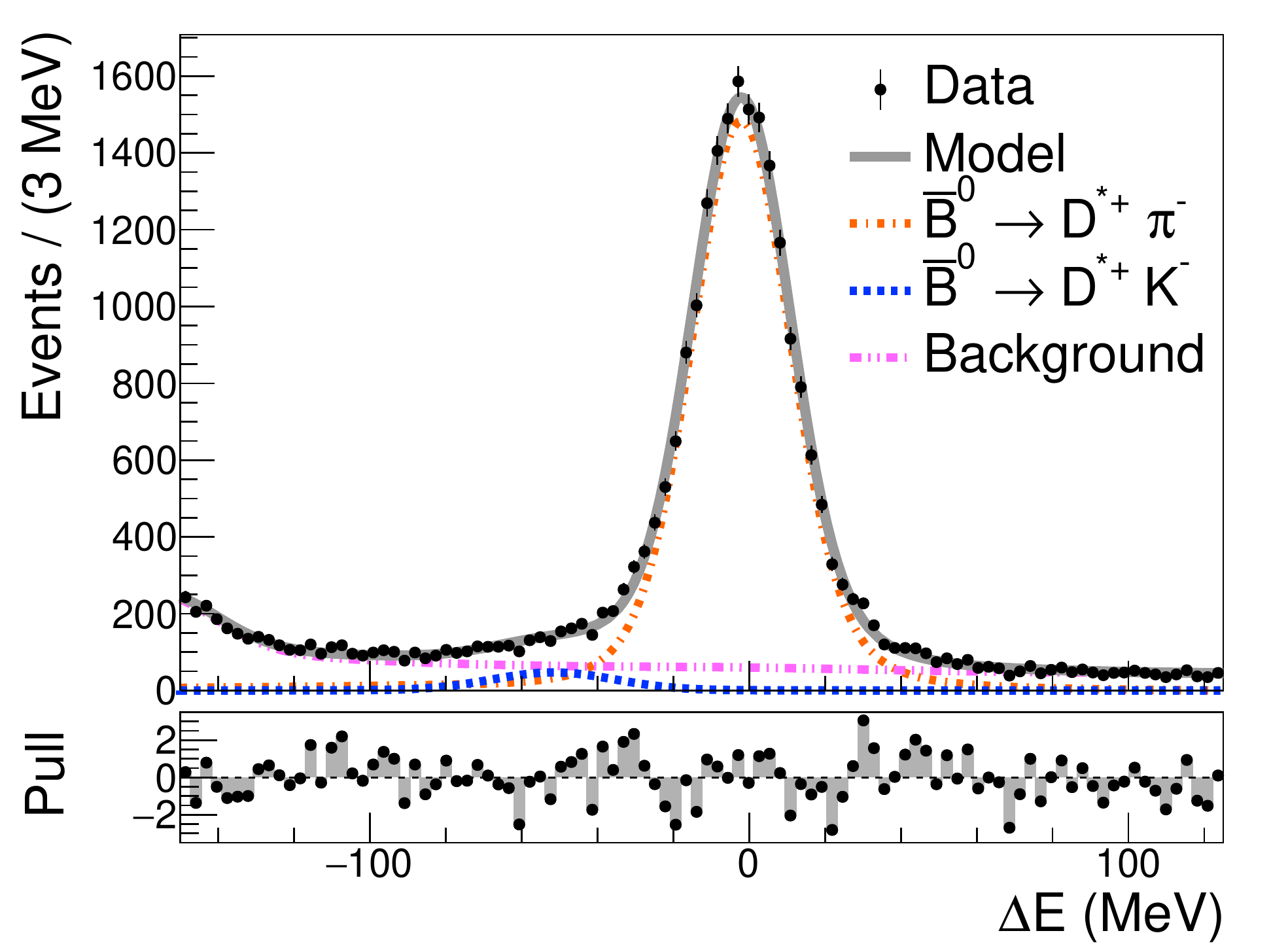}
		\caption{}
		\label{fig:final_results_fit_k3pi_pion}
	\end{subfigure}
	\caption{ 
		Results of the fits to the $\Delta E$ distributions in the $D^{0} \to K^- 2\pi^{+} \pi^{-}$ channel of (a) $\Bb^0 \to D^{*+} K^-$ and (b) $\Bb^{0} \to D^{*+} \pi^{-}$.  
	}
	\label{fig:final_fits_to_data_k3pi}
\end{figure*}

The yields obtained from the simultaneous fits are listed in Table~\ref{tab:yields} and the fits are shown in Figs.~\ref{fig:final_fits_to_data_kpi} and~\ref{fig:final_fits_to_data_k3pi} for the $D^0\to K^-\pi^+ $ and $D^0\to K^-2\pi^+ \pi^-$ channels, respectively. The yields for continuum processes and other $B$ meson decays were obtained from the fit separately and have been combined into a single background category for the table and figures.  
\begin{table*}[htb]
	\centering
	\caption{The signal and background event yields and their statistical uncertainties as obtained from the simultaneous fit, broken down by reconstruction channel.}
	\begin{tabular}{ c | c c c c }
		 Component & \multicolumn{2}{c}{$D^0\to K^- \pi^+$} & \multicolumn{2}{c}{$D^0\to K^- 2\pi^+ \pi^-$} \\
		 & $\Bb^0 \to D^{*+} \pi^-$ & $\Bb^0 \to D^{*+} K^-$ & $\Bb^0 \to D^{*+} \pi^-$ & $\Bb^0 \to D^{*+} K^-$\\
		 \hline
		 $\Bb^0 \to D^{*+} \pi^-$   & $16494 \pm 142$    & $1247 \pm 46$  & $19500 \pm 162$   & $1587 \pm 52$\\
		 $\Bb^0 \to D^{*+} K^-$     & $225 \pm 53$     & $1182 \pm 49$ & $731 \pm 71$      & $1414 \pm 55$\\
		 Background                     & $3390 \pm 115$    & $658 \pm 61$  & $7067 \pm 185$    & $1448 \pm 97$\\
	\end{tabular}\label{tab:yields}
\end{table*}


The branching fractions are then calculated from the measured signal yield of correctly identified candidates as
\begin{equation}
    \mathcal{B}(\Bb^0\to D^{*+} h^- ) = \frac{N_{{\rm meas}}(h^-) }{ N_{\Bb^0} \times \epsilon_h \times \mathcal{B}(D^{*+})\times \mathcal{B}(D^{0}) },
\end{equation}
where $\epsilon_h$ is the reconstruction efficiency for a given channel, derived from simulation and corrected for data-MC differences. The efficiency has an uncertainty due to the limited number of MC events that are used for its determination. The branching fractions $\mathcal{B}(D^{0}\to K^{-}2\pi^{+} \pi^-) =0.0822\pm 0.0014$, $\mathcal{B}(D^{0}\to K^{-}\pi^{+}) = 0.03946\pm 0.00030$ and $\mathcal{B}(D^{*+}\to D^{0}\pi^{+}) =0.667\pm0.005$ are taken from Ref. \cite{PDG}. The fraction of neutral \B meson decays is $ f({B^{0}}) = 0.486\pm 0.006$~\cite{PDG}; the total number of $\Bb^0$ mesons is given by
\begin{equation}
  N_{\Bb^0} = 2 \times N_{\BB}\times f({B^{0}}), 
\end{equation}
 where $ N_{\BB} = (771.6 \pm 10.6)\times 10^{6}$ is the total number of \BB meson pairs recorded at Belle.

\subsection{Systematic uncertainties}\label{sec:sys_errors}
There are five categories of systematic uncertainties: particle identification efficiencies, tracking efficiencies, PDFs, normalization parameters, and MC statistics.
We perform two types of measurements: branching fractions and ratios of branching fractions. For the latter, many sources of systematic uncertainty are fully correlated as the only difference between the two channels is the $K/\pi$ selection of the bachelor hadron. As a result, the correlated quantities cancel out and do not need to be considered in estimation of the systematic uncertainties. 

The first category of systematic uncertainty is based on $K/\pi$ identification corrections applied in bins of track polar angle and momentum, described in Sec.~\ref{sec:precuts}, which contain a statistical and systematic uncertainty. The uncertainty from $K/\pi$ identification corrections is calculated by varying the measured value by its uncertainty in each correction bin obtained with the calibration samples, taking into account correlations.

An uncertainty associated with the slow pion tracking efficiencies is evaluated based on corrections from a partially reconstructed $B^0\to D^{*-}\pi^+$ calibration sample binned in momentum, with a statistical uncertainty, and both bin-uncorrelated and bin-correlated systematic uncertainty components. The uncertainty from slow pion tracking efficiency is calculated by varying the measured value by its uncertainty in each correction bin obtained with the calibration sample, taking into account correlations. Track finding efficiencies for high-momentum tracks are assigned a flat systematic uncertainty of $0.35\%$ per track, derived from a partially reconstructed $D^{*+}$ calibration sample. 

Uncertainties arising due to PDF parameters determined from fits to MC simulation are considered, which include PDF fractions and shape parameters. The total uncertainty from this contribution is evaluated by varying each fixed parameter by one standard deviation and summing the uncertainties in quadrature. Fit biases are determined with pseudo-experiments (toy MC simulations) and the full estimated bias value is assigned as the uncertainty. 

The next categories of uncertainties are from normalization parameters, followed by MC statistics. For the branching fractions, the individual contributions are listed in Table~\ref{tab:sys_errs_brs}. 

For the ratio, only the values indicated with a dagger ($\dagger$) are considered, for $K/\pi$ selection these are calculated using only the bachelor hadrons. The total systematic uncertainty is found by summing these contributions in quadrature, under the assumption they are uncorrelated. The $D^0$ channels are combined by taking the average of measurements using the correlation coefficient, $\rho$, as given in Table~\ref{tab:sys_errs_brs}. There are three cases of possible correlation between related measurements: no correlation, partial or full correlation. For the tracking correlation coefficients we use the ratio of number of tracks: $N_{{\rm tracks}}(D^0\to K\pi)/N_{{\rm tracks}}(D^0\to K3\pi) = 3/5$.  The slow pion from the $D^{*+}$ decay is common to both channels and is treated separately due to its relatively large uncertainty. This uncertainty in slow pion detection efficiency is evaluated from a control sample that is statistically independent from the control sample for fast tracks. For the $\pi$-ID we use the ratio of the number of pions in the reconstructed decay channels, excluding the slow pion.

\begin{table*}[htb]
\caption{
Breakdown of the statistical and systematic uncertainties (in $\%$). The total is determined assuming zero correlations between the individual uncertainties. The entries marked with a $\dagger$ propagate into the ratio while others cancel out. The correlation coefficient used to combine the $D^0$ channels is denoted as $\rho$. In the $\rho$ column the first value for the $\pi$-ID systematic uncertainties is for the $\Bb \to D^{*+} \pi^-$ combination and the second is for the $\Bb \to D^{*+} K^-$ combination.}
\begin{tabular}{c | c c  c c  c c |c}  
	&  \multicolumn{2}{c}{ $\boldsymbol{{D^0\to K^-\pi^+}}$} & \multicolumn{2}{c}{ $\boldsymbol{{D^0\to K^-2\pi^+\pi^-}}$} &\multicolumn{2}{c}{\textbf{Combined} } & \\
	type	& $\Bb \to D^{*+} \pi^-$        & $\Bb \to D^{*+} K^-$ &$\Bb \to D^{*+} \pi^-$   & $\Bb \to D^{*+} K^-$  & $\Bb \to D^{*+} \pi^-$   & $\Bb \to D^{*+} K^-$ & $\rho$ \\ \hline
\multirow{2}{*}{$\pi$-ID stat.  } & $0.78$ & \multirow{2}{*}{$0.54$}  & $0.95$& \multirow{2}{*}{$0.20$} & $0.75$         & \multirow{2}{*}{$0.32$} & \multirow{2}{*}{2/4, 1/3} \\
& $0.72^\dagger$& & $0.65^\dagger$ &             & $0.58^\dagger$ &                           &                      \\[0.2cm]
\multirow{2}{*}{$\pi$-ID sys.}    & $0.60$        & \multirow{2}{*}{$0.27$}  & $0.52$                  & \multirow{2}{*}{$0.20$} & $0.49$         & \multirow{2}{*}{$0.19$} & \multirow{2}{*}{2/4, 1/3} \\
                                  & $0.44^\dagger$&                            & $0.46^\dagger$          &                           & $0.41^\dagger$ &                           &                      \\[0.2cm]
\multirow{2}{*}{$K$-ID stat.}     & \multirow{2}{*}{$0.76$} & $1.05$        & \multirow{2}{*}{$0.72$} & $1.03$              & \multirow{2}{*}{$0.74$} & $1.04$         & \multirow{2}{*}{1} \\
                                  &                           & $0.72^\dagger$&                           & $0.72^\dagger$      &                           & $0.64^\dagger$ &                      \\[0.2cm]
\multirow{2}{*}{$K$-ID sys.}      & \multirow{2}{*}{$0.53$} & $1.15 $         & \multirow{2}{*}{$0.57$} & $0.62$            & \multirow{2}{*}{$0.55$} & $0.89$         & \multirow{2}{*}{1} \\
                                  &                           & $0.61^\dagger$  &                           & $0.62^\dagger$    &                           & $0.55^\dagger$ &                      \\ [0.2cm]
$K$-ID run dep. sys.              & $0.30$ & $0.30$                   & $0.30$ & $0.30$               & $0.30$ & $0.30$ & 1  \\ \hline
Slow $\pi$ stat.          & $0.79$ & $0.79$                   & $0.79$ & $0.79$               & $0.79$ & $0.79$ & 1\\
Slow $\pi$ sys.           & $0.01$ & $0.01$                   & $0.01$ & $0.01$               & $0.01$ & $0.01$ & 1\\
Slow $\pi$ corr.          & $1.33$ & $1.33$                   & $1.33$ & $1.33$               & $1.33$ & $1.33$ & 1\\
           Tracking sys.          & $1.05$ & $1.05$                   & $1.75$ & $1.75$               & $1.26$ & $1.26$ & 3/5 \\ \hline

Bkg. PDF: fixed yields             & $0.10^\dagger$  & $0.10^\dagger$  & $0.10^\dagger$ & $0.10^\dagger$ & $0.07^\dagger$ & $0.07^\dagger$ & 0\\  
Bkg. PDF: fixed shapes             & $0.10^\dagger$  & $0.10^\dagger$  & $0.10^\dagger$ & $0.10^\dagger$ & $0.07^\dagger$ & $0.07^\dagger$ & 0\\ 
Fit bias                          & $0.15^\dagger$   & $0.15^\dagger$   & $0.08^\dagger$ & $0.74^\dagger$ & $0.09^\dagger$ & $0.37^\dagger$ & 0 \\ \hline
$N_{\Bb^0}$                    & $1.84$ & $1.84$                   & $1.84$ & $1.84$                 & $1.84$ & $1.84$ & 1 \\ 
$\mathcal{B}(D^{*+}\to D^0\pi^+)$ & $0.74$ & $0.74$                   & $0.74$ & $0.74$                 & $0.74$ & $0.74$ & 1 \\  
$\mathcal{B}(D^0)$                & $0.78$ & $0.78$                   & $1.70$ & $1.70$                 & $0.94$ & $0.94$ & 0 \\  \hline
MC stat.                          & $0.39 ^\dagger$ & $1.40 ^\dagger$  & $0.35^\dagger$ & $1.39^\dagger$ & $0.26^\dagger$ & $0.99^\dagger$ & 0\\ \hline
Total sys. ($\mathcal{B}$)             & $3.20$ & $3.60$                   & $3.82$ & $4.06$                 & $3.26$ & $3.47$ &   \\ 
Total sys. (ratio)                     & $1.93$ & $1.93$                   & $1.89$ & $1.89$                 & $1.50$ & $1.50$ &   \\ \hline
Total stat.                    & $0.84$ & $4.00$                   & $0.78$ & $3.70$                 & $0.57$ & $2.74$ &   \\
		\end{tabular}
		\label{tab:sys_errs_brs}
\end{table*}

\subsection{Branching fraction results}\label{sec:results}
\begin{figure*}[htb]
	\centering
	\begin{subfigure}{0.45\textwidth}
		\centering
		\includegraphics[width=\textwidth]{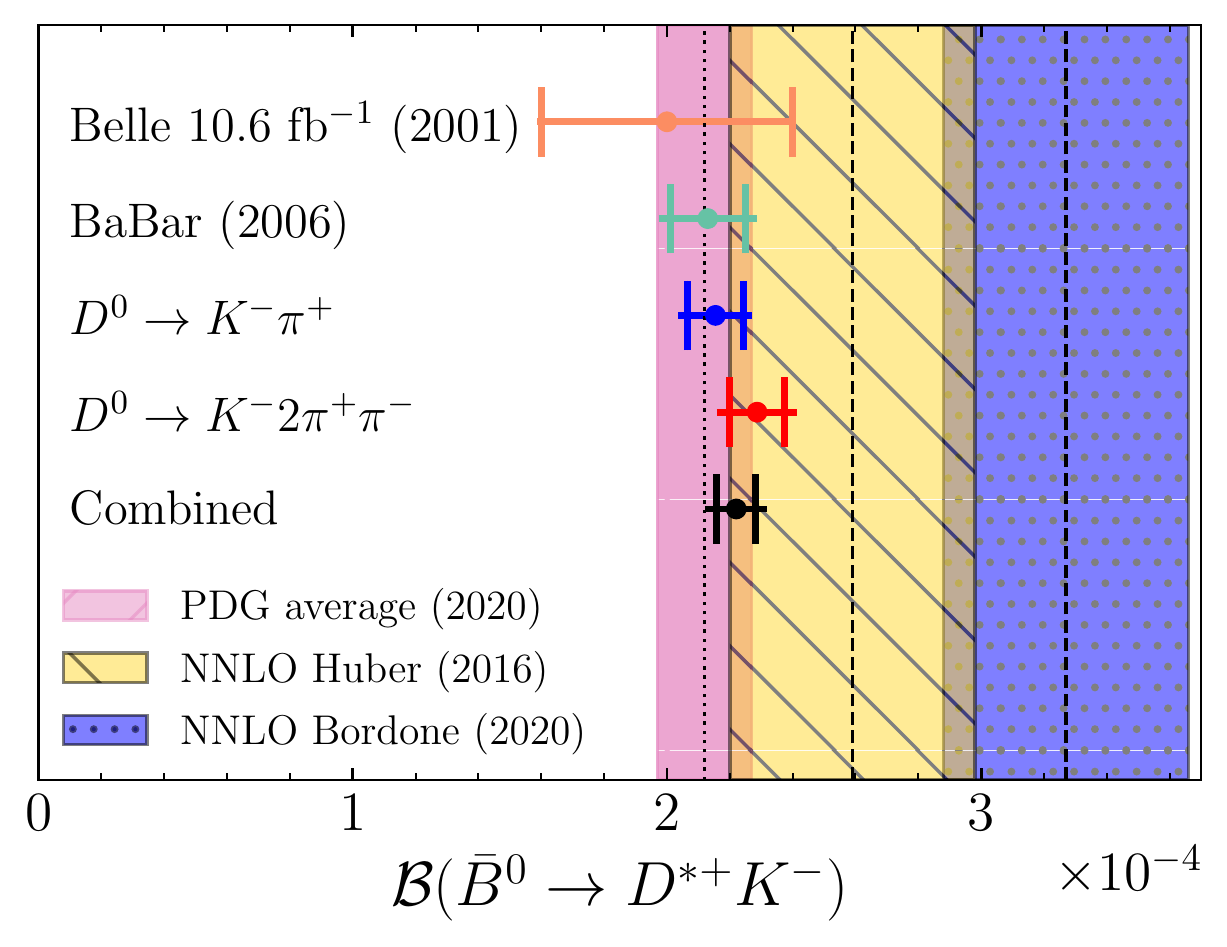}
		\caption{}
		\label{fig:final_results_br_kaon}
	\end{subfigure}
	\begin{subfigure}{0.45\textwidth}
		\centering
		\includegraphics[width=\textwidth]{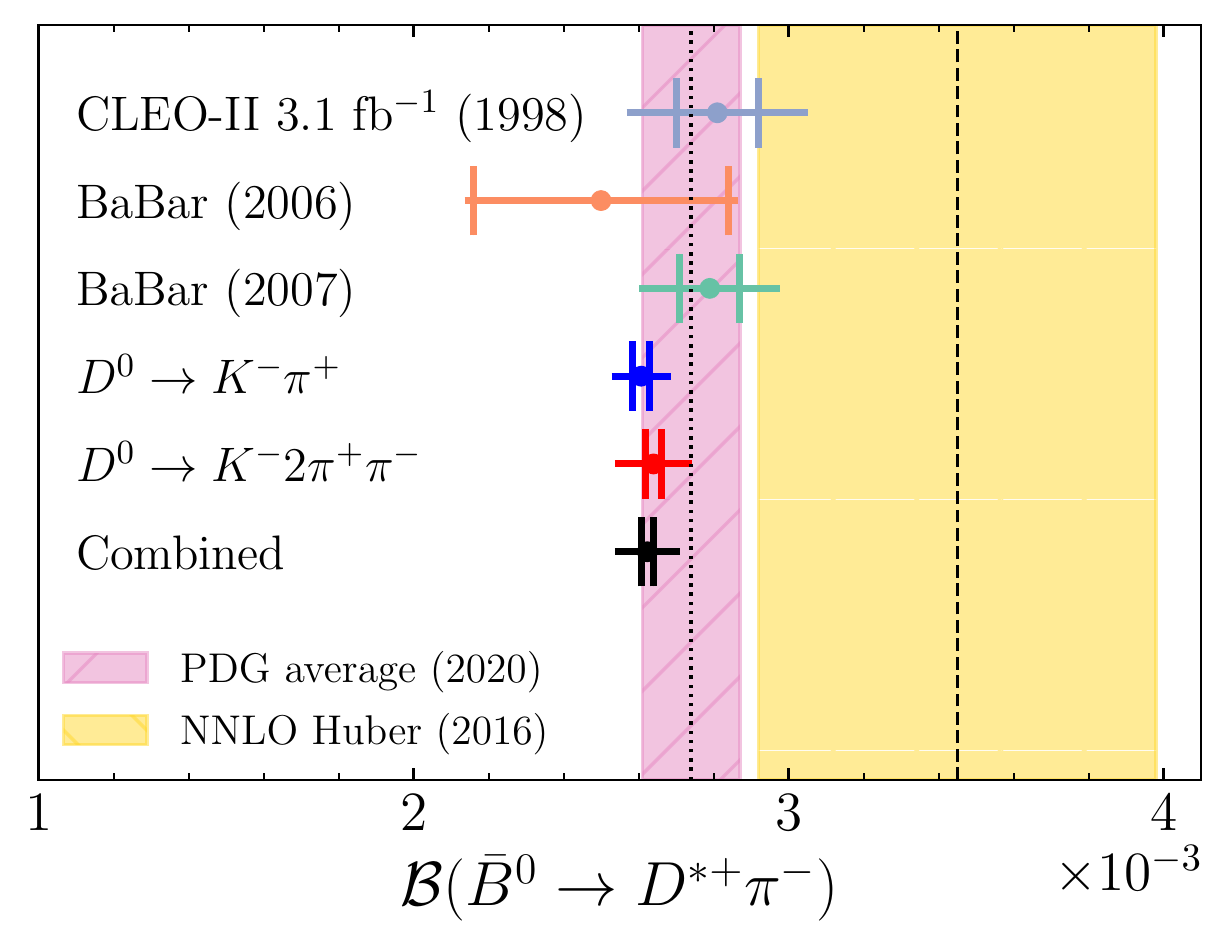}
		\caption{}
		\label{fig:final_results_br_pion}
	\end{subfigure}
	\caption{ 
		Comparison of the branching fraction ratio measurements using the full data sample and the data subsamples with respect to previous measurements by (a) Belle~\cite{Abe:2001waa} and BaBar~\cite{Aubert:2005yt}, and (b) BaBar~\cite{Aubert:2006jc, BaBar:2006rof} and CLEO-II~\cite{CLEO:1997vmd}. The theoretical predictions are taken from Refs.~\cite{Huber:2016xod,Bordone:2020gao}. The inner uncertainty is statistical and the outer is the quadrature sum of both statistical and systematic uncertainties.
	}
	\label{fig:final_results_br}
\end{figure*}

\begin{figure}[htb]
	\centering
	\begin{subfigure}{0.45\textwidth}
		\centering
		\includegraphics[width=\textwidth]{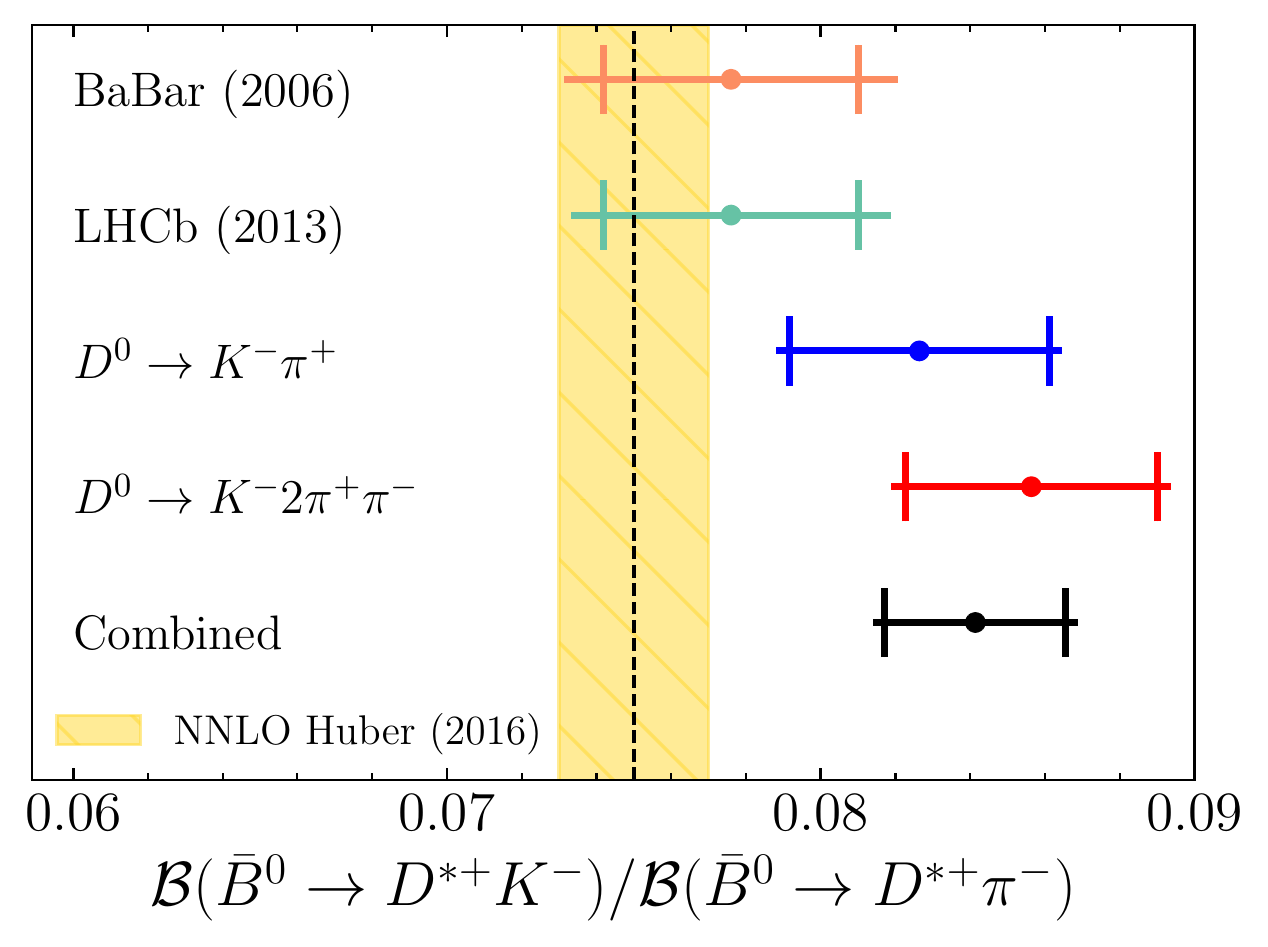}
	\end{subfigure}
	\caption{ 
		Comparison of the branching fraction ratio measurements using the full data sample and the data subsamples with respect to previous measurements by BaBar~\cite{Aubert:2005yt}, LHCb~\cite{Aaij:2013xca} and the theoretical prediction from Ref.~\cite{Huber:2016xod}. The inner uncertainty is statistical and the outer is the quadrature sum of both statistical and systematic uncertainties.
	}
	\label{fig:final_results_ratio}
\end{figure}

The branching fractions and their ratios are shown in Figs.~\ref{fig:final_results_br} and~\ref{fig:final_results_ratio}, respectively, with a comparison to theoretical predictions and prior measurements. The numerical values are listed in Table~\ref{tab:results_br}. For future updates of the $D^{*+}$ and $D^{0}$ meson branching fractions we give results for the products $\mathcal{B}(\Bb^0\to D^{*+} h^- )\times \mathcal{B}(D^{*+})\times \mathcal{B}(D^{0}) = N_{{\rm meas}}(h^-) / ( N_{\Bb^0} \times \epsilon_h )$ in Table~\ref{tab:renormalise_results}.

For $\mathcal{B}(\Bb^0 \to D^{*+} K^-) $ the results are compatible with the previous Belle measurement performed on a $10.4\,{\rm fb}^{-1}$ ($N_{\BB} = 11.1\times 10^6$) dataset \cite{Abe:2001waa}. Both the statistical and systematic uncertainties improved due to a larger dataset and better understanding of the detector. The values are compared to two theory models, and when taking uncertainties from experiment and theory into account, there is a $1.0 \sigma $ deviation from the predictions of Huber et al. at Next-to-Next-to-Leading-Order (NNLO)~\cite{Huber:2016xod} and a deviation of $2.7\sigma$ with respect to Bordone et al.~\cite{Bordone:2020gao}. 
The same evaluation is made for $\mathcal{B}(\Bb^0 \to D^{*+} \pi^-) $ with a deviation of $1.7\sigma$ with respect to Ref.~\cite{Huber:2016xod}.
For the ratio $\mathcal{R}_{K/\pi}=\mathcal{B}(\Bb^0 \to D^{*+} K^-)/\mathcal{B}(\Bb^0 \to D^{*+} \pi^-)$ a deviation of $2.7\sigma$ from Ref.~\cite{Huber:2016xod} is found. The total experimental uncertainty on this ratio is 3.2\%, which is lower than BaBar ($5.7\%$)~\cite{Aubert:2005yt} and LHCb ($5.5\%$)~\cite{Aaij:2013xca}.

\begin{table*}\caption {
Results of the branching fraction and their ratios. The first uncertainty is statistical and the second is systematic. The last column lists the deviation with theoretical predictions in terms of standard deviations, $\sigma$, taking into account experimental and theoretical uncertainties.
The comparisons without parentheses are with respect to Huber et al.~\cite{Huber:2016xod}, and those with parentheses are with respect to Bordone et al.~\cite{Bordone:2020gao}. 
}
	\centering
	\begin{tabular}{c | c c} 
		$\mathcal{B}(\Bb^0 \to D^{*+} \pi^-) $  & Result &  $n\sigma$ meas.$-$theo. \\
		\hline
		$D^0\to K^- \pi^+ $                         & $\brpionkpi$ & $1.8$ \\
		$D^0\to K^- 2\pi^+ \pi^- $                  &  $\brpionkthreepi$ & $1.7$ \\
		Combined                                    & $\brpioncomb$ & $1.7 $ \\
		
		\addlinespace[1.5ex] 		$\mathcal{B}(\Bb^0 \to D^{*+} K^-) $ & & \\
		\hline
		$D^0\to K^- \pi^+ $                         &  $\brkaonkpi$&  $1.1~(2.7)$ \\
		$D^0\to K^- 2\pi^+ \pi^- $                  &   $\brkaonkthreepi$ & $0.7~(2.4)$ \\		 
		Combined                                    &  $\brkaoncomb$& $0.9~(2.6)$ \\
		
		\addlinespace[1.5ex] 	$\mathcal{R}_{K/\pi}$ & & \\
				\hline
		$D^0\to K^- \pi^+ $ &  $\ratiokpi$ & $1.8$  \\
		$D^0\to K^- 2\pi^+ \pi^- $ & $\ratiokthreepi$  & $2.5$\\
		Combined & $\ratiocomb$& $2.7$ \\
	\end{tabular}
	\label{tab:results_br}
\end{table*}

\begin{table*}[htb]
	\centering
	\caption{Results for $\mathcal{B}(\Bb^0\to D^{*+} h^- )\times \mathcal{B}(D^{*+})\times \mathcal{B}(D^{0}) = N_{{\rm meas}}(h^-) / ( N_{\Bb^0} \times \epsilon_h )$. The first uncertainty value is statistical and the second is systematic.}
	\begin{tabular}{ c | c | c }
		 & $D^0\to K^- \pi^+$ & $D^0\to K^- 2\pi^+ \pi^-$ \\ \hline
		 $B^0\to D^{*+} \pi^-$&$(6.862\pm0.058\pm 0.206 ) \times 10^{-5}$ & $(1.509\pm0.012\pm0.046)\times 10^{-4}$\\
		 $B^0\to D^{*+} K^-$  & $(5.671\pm0.227\pm0.195)\times 10^{-6}$ & $(1.307\pm0.048\pm0.040)\times 10^{-5}$ \\
	\end{tabular}\label{tab:renormalise_results}
\end{table*}

\section{Measurement of $\left|a_1(h)\right|$ }
Quantum Chromodynamic (QCD) factorization predicts $|a_1(h)|=1$ in its most naive version. Taking higher order corrections into account one expects a quasi-universal value of $|a_1(h)|=1.05$~\cite{Beneke:2000_qcd_factorisation}, independent of the bachelor hadron species. 

\begin{table*}[htb]
	\centering
	\caption{Input parameters for the $|a_1(h)|$ calculation taken from Ref.~\cite{PDG}. Values used exclusively in the determination of the semileptonic decay rates not listed here are taken from Ref.~\cite{Ferlewicz:2020lxm}.}
	\begin{tabular}{ c c  c  }
		Description & Parameter & Value \\
		\hline 
		Lifetime &$\tau_{B^0}$ & $(1.519\pm0.004)$ ps\\
	    CKM matrix element & $|{V_{ud}}|$ & $0.97370\pm0.00010$ \\
		&$|{V_{us}}|$ & $0.0.2231\pm0.0004$ \\
		Decay constants&$f_{\pi}$ & $(0.1302\pm0.0012)$ \gevcc \\
		&$f_{K}$ & $(0.1556\pm0.0004)$ \gevcc \\
		&$f_{\rho}$ & $(0.216\pm0.006 )$ \gevcc \\
		&$f_{K^*}$ & $(0.211\pm0.007)$ \gevcc \\
		&$f_{a_1}$ & $(0.238\pm0.01)$ \gevcc \\
		&$f_{D}$ & $(0.2119\pm0.0011)$ \gevcc \\
		&$X_h$ & $1\pm0.0007$ \\
		Branching fraction&$\mathcal{B}(\Bb\to D^{*+}\rho^- )$ & $(6.8\pm0.9)\times 10^{-3} $ \\
		&$\mathcal{B}(\Bb\to D^{*+}K^{*-} )$ & $(3.3\pm0.6)\times 10^{-4} $ \\
		&$\mathcal{B}(\Bb\to D^{*+}a_1^{-} )$ & $(1.30\pm0.27)\times 10^{-2} $ \\
		&$\mathcal{B}(D^{*+}\to D^{0}\pi )$ & $0.667\pm 0.005 $ \\
		&$\mathcal{B}(D^{0}\to K^{-}\pi^{+} )$ & $(3.946\pm 0.030)\times 10^{-2}$ \\
		&$\mathcal{B}(D^{0}\to K^{-}2\pi^{+}\pi^{-} )$ & $(8.22\pm 0.14)\times 10^{-2}$ \\
		Masses&$m(\pi^+)$ & $(0.13957039\pm0.00000018) $ \gevcc \\
		&$m_{K^+}$ & $(0.493677\pm0.000016) $ \gevcc \\
		&$m_{\rho^+}$ & $(0.77526\pm0.00025) $ \gevcc \\
		&$m_{K^{*+}}$  & $(0.89167\pm0.00026) $ \gevcc \\
		&$m_{a_1^{+}}$ & $(1.230\pm0.040) $ \gevcc \\
		\hline
	\end{tabular}\label{tab:inputs}
\end{table*}

\begin{figure}[htb]
	\centering
	\begin{subfigure}{0.45\textwidth}
		\centering
		\includegraphics[width=\textwidth]{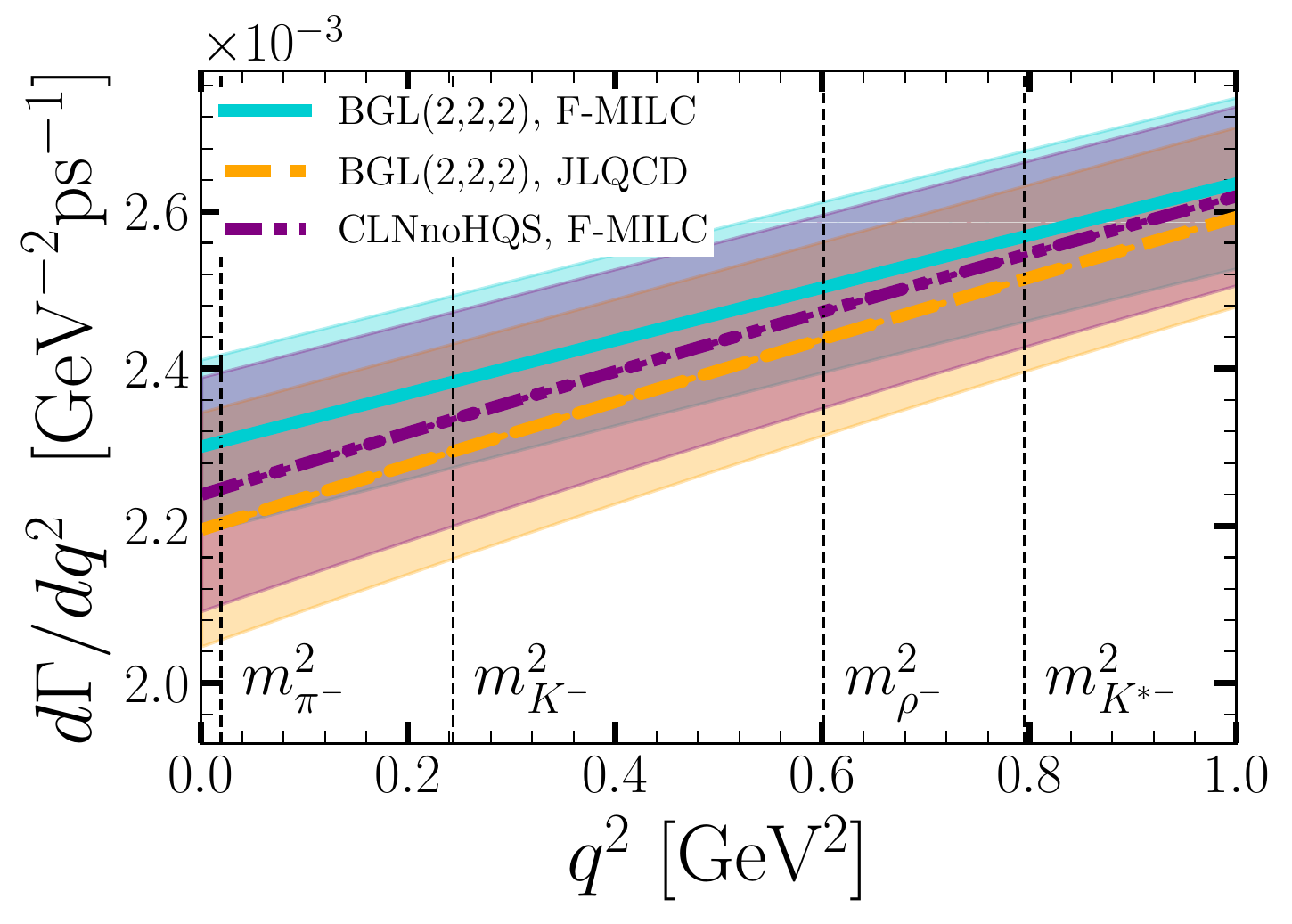}
	\end{subfigure}
	\caption{Semileptonic decay rates $d\Gamma(\Bb^0 \to D^{*+}  \ell^- \nub ) /dq^2$ as a function of dilepton invariant mass squared determined from fits to Belle data and lattice QCD inputs from Fermilab-MILC and JLQCD, based on the method described in Ref.~\cite{Ferlewicz:2020lxm}.}
	\label{fig:q2_results}
\end{figure}

The values for the differential decay rate $d\Gamma(\Bb^0 \to D^{*+}  \ell^- \nub ) /dq^2$ are directly extracted from an untagged Belle measurement~\cite{Waheed_2019}.
The semileptonic differential decay rate is determined by fits using both the Caprini-Lellouch-Neubert (CLN)~\cite{Caprini:1997mu} and Boyd-Grinstein-Lebed (BGL)~\cite{Boyd_1997} parameterizations using additional constraints from Lattice QCD (LQCD) calculations of form factors at nonzero recoil, as described in Ref.~\cite{Ferlewicz:2020lxm}. Two independent sources of LQCD calculations were included so that any dependence on the inputs could be tested. These inputs were from the Fermilab-MILC collaboration \cite{FermilabLattice:2021cdg}, using nine different values, and JLQCD \cite{Kaneko:2020ss} with four. A comparison of the differential decay rates is shown in Fig.~\ref{fig:q2_results} for the CLN noHQS configuration using JLQCD inputs and BGL(2,2,2) in both JLQCD and Fermilab-MILC. Each of the differential decay rates is consistent within the uncertainty bands. 
The BGL(2,2,2) configuration was taken as the nominal result as it is more model-independent than CLN, and lattice inputs from Fermilab-MILC were included due to more values being available with a robust uncertainty estimation. Further inputs used in the evaluation of $|a_1(h)|$ (Eq.~\ref{eq:diff_decay_rate}) are listed in Table~\ref{tab:inputs}, and are taken from Ref.~\cite{PDG}. The parameter $X_{h}$ depends on the spin of $h$: $X_{h}=1$ for vector mesons and $X_{h}=1 \pm m_h^2/m_B^2 $ for non-vector mesons. 

\subsection{Testing $SU(3)$ symmetry}
A test is performed to verify whether $|a_1(h)|$ is a universal factor independent of the quarks involved in the hadronic transition. A value of $|a_1(K)|^2/|a_1(\pi)|^2=1$ would imply that $SU(3)$ symmetry holds, as suggested in Ref.~\cite{Beneke:2000_qcd_factorisation}. The test is done by measuring the ratios of $|a_1(h)|$ for different particles species, i.e. $K$ and $\pi$:

\begin{equation}
\begin{split}
	\frac {|a_1(K)|^2}{|a_1(\pi)|^2} =
	& \frac{|V_{ud}|^2}{|V_{us}|^2} \frac{f_{\pi}^2}{f_{K}^2} \frac{X_{\pi}}{X_{K}} R_{K/\pi}\\
	& \left( \frac{ \  d\Gamma(\Bb^0 \to D^{*+}  \ell^- \nub ) / dq^2 |_{q^2=m^2_\pi} } { d\Gamma(\Bb^0 \to D^{*+}  \ell^- \nub ) /dq^2 |_{q^2=m^2_K}  } \right).
\end{split}
\end{equation}

Two sets of ratios are performed: ratios based on hadronic branching fractions measured in this paper ($h=K$, $\pi$), and ratios based on branching fractions listed in Ref.~\cite{PDG} ($h=\rho$, $K^*$, $a_1$)\footnote{Where $a_1$ is written as a function of $q^2$ or $h$ it refers to the QCD factorization parameter, and when it is written alone it refers to the meson $a_1^+(1260)$.}.

\subsection{Systematic uncertainties}

For the ratios calculated with Belle data ($h=K$, $\pi$) many systematic uncertainties are fully correlated and cancel out in the evaluation of $|a_1(h)|$ and their ratios. Furthermore, it was verified that external input parameters match, in particular $\mathcal{B}(D^{*+})$ and $\mathcal{B}(D^0)$. This means that for the $\Bb^0 \to D^{*+} h^{-} $ decay widths, only the $K/\pi$ selection of the bachelor hadron, fit PDF parameters, fit bias, and MC statistical uncertainty are considered. For the $\Bb^0 \to D^{*+} \ell^{-} \nub$ differential decay rate we consider PDF related uncertainties, statistical uncertainties, as well as lepton identification, fake $e/\mu$ rates and the $D^{**}$ branching fractions and form factors as sources of systematic uncertainties. The numerical values can be found in Ref.~\cite{Waheed_2019}.
For all calculations featuring the mesons $h=\rho$, $K^*$, $a_1$, the full systematic uncertainty is taken.
For ratios of $|a_1(h_1)|^2/|a_1(h_2)|^2$, where $h_1$ and $h_2$ are different particle types, it is crucial to also account for the correlation between different $q^2$ points in the $\Bb^0 \to D^{*+} \ell^{-} \nub$ differential decay rate spectrum. These correlations are found from the toy MC samples provided by the authors of Ref.~\cite{Ferlewicz:2020lxm} and are listed in Table~\ref{tab:q2_correlations}. We also note that the hadronic recoil binning for the semileptonic differential decay rate has both $q^2 = m^2_{\pi^-}$ and $m^2_{K^-}$ contained within the same bin, resulting in a full correlation between their corresponding $|a_1(h)|$ values. A breakdown of the relative uncertainty contributions for the $|a_1(h)|$ and $|a_1(h_1)|^2/|a_1(h_2)|^2$ measurements for pions and kaons is given in Table~\ref{tab:err_breakdown}.

\begin{table}[htb]
	\centering
	\caption{Correlations of $d\Gamma (\Bb^0 \to D^{*+} \ell^{-} \nub)/dq^2$  between different $q^2 = m_{{h}}^2 $ points.}
	\begin{tabular}{ c | c c c c c }
		
    BGL F-MILC &$\pi^{-}$&     $K^{-}$&  $\rho^{-}$  &  $K^{*-}$&  $a_1^{-}$ \\
\hline
$\pi^{-}$  & $1.000$ & $0.996$ & $0.977$ & $0.960$  & $0.874$ \\
$K^{-}$     &  & $1.000$ & $0.991$ & $0.980$  & $0.911$ \\
$\rho^{-}$  &  &  & $1.000$ & $0.998$  & $0.957$ \\
$K^{*-}$    &  &  &  & $1.000$  & $0.974$ \\
$a_1^{-} $&  &  &  &   & $1.000$ \\
\addlinespace[1.5ex] 
    BGL JLQCD &$\pi^{-}$&     $K^{-}$&  $\rho^{-}$  &  $K^{*-}$&  $a_1^{-}$ \\
\hline
$\pi^{-}$   & $1.000$ & $0.994$ & $0.956$ & $0.918$  & $0.713$\\
$K^{-}$     &  & $1.000$ & $0.982$ & $0.956$  & $0.783$\\
$\rho^{-}$  &  &  & $1.000$ & $0.994$  & $0.886$\\
$K^{*-}$    &  &  &  & $1.000$  & $0.931$\\
$a_1^{-}$ &  &  &  &   & $1.000$\\
\addlinespace[1.5ex] 
CLN JLQCD &$\pi^{-}$&     $K^{-}$&  $\rho^{-}$  &  $K^{*-}$&  $a_1^{-}$ \\
\hline
$\pi^{-}$   & $1.000$ & $0.992$ & $0.941$ & $0.891$  & $0.623$\\
$K^{-}$     &  & $1.000$ & $0.976$ & $0.940$  & $0.713$\\
$\rho^{-}$  &  &  & $1.000$ & $0.992$  & $0.848$\\
$K^{*-}$    &  &  &  & $1.000$  & $0.909$\\
$a_1^{-}$ &  &  &  &   & $1.000$\\
	\end{tabular}\label{tab:q2_correlations}
\end{table}

\begin{table*}[htb]
    \centering
    \caption{
    Breakdown of the contributions to the total uncertainty on the $|a_1(h)|$ and $|a_1(h_1)|^2/|a_1(h_2)|^2$ measurements, given as a relative percentage. The hadronic uncertainties estimated from the $\Bb^0 \to D^{*+} h^-$ analysis are separated into statistical and systematic categories, otherwise they are combined. The `Other' category combines all uncertainties from Standard Model constants.
    }
    
    \begin{tabular}{c|cc|ccc|c}
         & \multicolumn{2}{c|}{Hadronic} & \multicolumn{3}{c|}{Combined Semileptonic} & \\ 
        Measurement & stat. & sys. & BGL, F-MILC & BGL, JLQCD & CLN, JLQCD & Other\\
        \hline  
$|a_1(\pi)|$                          & 0.4 & 0.3                 & 1.5       & 2.1  & 2.0    & 1.1 \\
$|a_1(K)|$                            & 2.1  & 0.9                & 1.4       & 1.8  & 1.8    & 0.4 \\
$|a_1(\rho)|$                         & \multicolumn{2}{c|}{6.6}  & 1.3       & 1.5  & 1.4    & 2.8 \\
$|a_1(K^{*})|$                         & \multicolumn{2}{c|}{9.1}  & 1.3       & 1.4  & 1.3    & 3.3 \\
$|a_1(a_1)|$                      & \multicolumn{2}{c|}{10.4} & 1.2       & 1.1  & 1.1    & 4.2 \\
$|a_1(K)|^2/|a_1(\pi)|^2$       & 4.2  & 1.8                & 0.3       & 0.7  & 0.7    & 2.2 \\
$|a_1(\rho)|^2/|a_1(\pi)|^2$    & \multicolumn{2}{c|}{13.3} & 0.7       & 1.5  & 1.7    & 6.0 \\
$|a_1(K^{*})|^2/|a_1(\pi)|^2$      & \multicolumn{2}{c|}{18.2} & 0.9       & 1.9  & 2.1    & 7.0 \\
$|a_1(a_1)|^2/|a_1(\pi)|^2$   & \multicolumn{2}{c|}{20.8} & 1.5       & 3.0  & 3.2    & 8.7 \\
$|a_1(\rho)|^2/|a_1(K)|^2$      & \multicolumn{2}{c|}{14.0} & 0.4       & 0.9  & 1.0    & 5.6 \\
$|a_1(K^{*})|^2/|a_1(K)|^2$        & \multicolumn{2}{c|}{18.7} & 0.6       & 1.3  & 1.4    & 6.7 \\
$|a_1(a_1)|^2/|a_1(K)|^2$     & \multicolumn{2}{c|}{21.2} & 1.2       & 2.3  & 2.5    & 8.4 \\
$|a_1(K^{*})|^2/|a_1(\rho)|^2$     & \multicolumn{2}{c|}{22.5} & 0.2       & 0.4  & 0.4    & 8.7 \\
$|a_1(a_1)|^2/|a_1(\rho)|^2$  & \multicolumn{2}{c|}{24.6} & 0.8       & 1.4  & 1.5    & 10.1 \\
$|a_1(a_1)|^2/|a_1(K^{*})|^2$    & \multicolumn{2}{c|}{27.6} & 0.6       & 1.0  & 1.1    & 10.7 \\
    \end{tabular}\label{tab:err_breakdown}

\end{table*}

\subsection{Results for $\left|a_1(h)\right|$}
\begin{figure}[htb]
	\centering
	\begin{subfigure}{.45\textwidth}
		\centering
		\includegraphics[width=\textwidth]{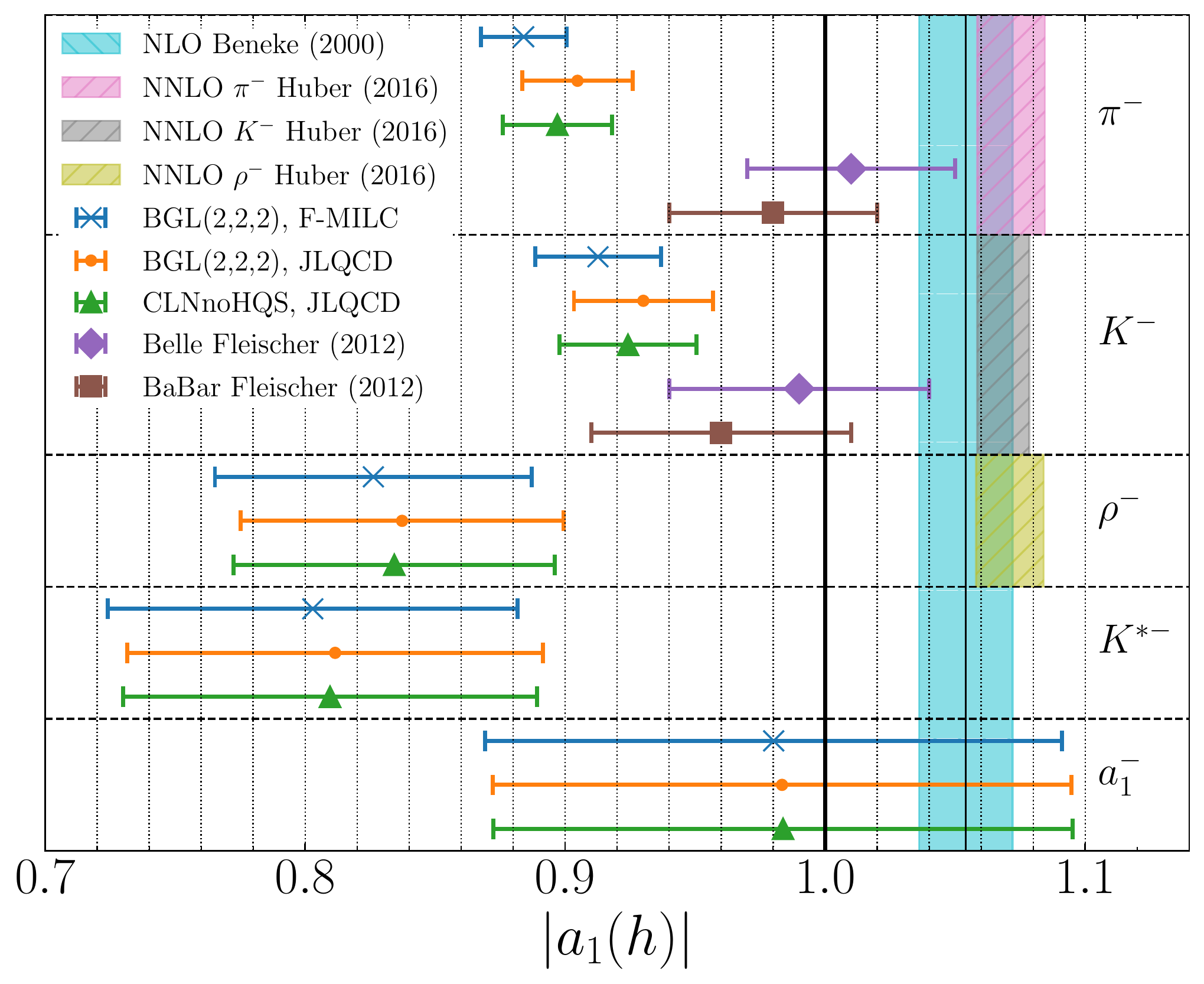}
	\end{subfigure}
	\caption{The extracted values of $|a_1(h)|$ from $\bar B^0 \to D^{*+}h^-$, $h=\pi,~K, ~\rho,~K^*,~a_1$ using the three semileptonic input scenarios described in the text. The theory predictions are taken from Refs.~\cite{Beneke:2000_qcd_factorisation, Huber:2016xod}. }
	\label{fig:a1-results}
\end{figure}

The results for $|a_1(h)|$ are given in Table~\ref{tab:a1_results}, and compared to theoretical prediction and previous evaluations in Fig.~\ref{fig:a1-results}. For $|a_1(\rho)|$, $|a_1(K^{*})|$ and $|a_1(a_1)|$ the hadronic \B decay branching fractions are taken from Ref.~\cite{PDG}.
\begin{table*}[htb]
	\centering
	\caption{The extracted values of $|a_1(h)|$ for $h=\pi,~\rho,~K^*,~a_1$, and the three semileptonic input scenarios described in the text. The deviations are calculated with respect to predictions in Ref.~\cite{Huber:2016xod} and take both the experimental and theoretical uncertainties into account.}
	\begin{tabular}{c | c | c | c  c }
	Particle & Model & $|a_1(h)|$ & $n\sigma$ (meas.-theo.)\\
\hline
$\pi^{-}$& BGL(2,2,2), F-MILC & $0.884 \pm 0.016$ & $8.9$ $(-18\pm2)\%$ \\  
& BGL(2,2,2), JLQCD & $0.905 \pm 0.021$ & $6.7$ $(-16\pm2)\%$ \\  
& CLNnoHQS, JLQCD & $0.897 \pm 0.021$ & $7.1$ $(-16\pm2)\%$ \\  
\hline
$K^{-}$& BGL(2,2,2), F-MILC & $0.913 \pm 0.024$ & $5.8$ $(-15\pm3)\%$ \\  
& BGL(2,2,2), JLQCD & $0.930 \pm 0.027$ & $4.7$ $(-13\pm3)\%$ \\  
& CLNnoHQS, JLQCD & $0.924 \pm 0.026$ & $5.0$ $(-14\pm3)\%$ \\  
\hline
$\rho^{-}$& BGL(2,2,2), F-MILC & $0.826 \pm 0.061$ & $3.6$ $(-22\pm5)\%$ \\  
& BGL(2,2,2), JLQCD & $0.837 \pm 0.062$ & $3.4$ $(-21\pm6)\%$ \\  
& CLNnoHQS, JLQCD & $0.834 \pm 0.062$ & $3.4$ $(-21\pm6)\%$ \\  
\hline
$K^{*-}$& BGL(2,2,2), F-MILC & $0.803 \pm 0.079$ & $3.1$ $(-24\pm8)\%$ \\  
& BGL(2,2,2), JLQCD & $0.812 \pm 0.080$ & $3.0$ $(-23\pm8)\%$ \\  
& CLNnoHQS, JLQCD & $0.810 \pm 0.080$ & $3.0$ $(-23\pm8)\%$ \\  
\hline
$a_1^{-}$& BGL(2,2,2), F-MILC & $0.980 \pm 0.111$ & $0.7$ $(-7\pm11)\%$ \\  
& BGL(2,2,2), JLQCD & $0.983 \pm 0.111$ & $0.6$ $(-7\pm11)\%$ \\  
& CLNnoHQS, JLQCD & $0.984 \pm 0.111$ & $0.6$ $(-7\pm11)\%$ \\
	\end{tabular}\label{tab:a1_results}
\end{table*}

Nominal values of $|a_1(\pi)|=\aOnePion$ from $\Bb\to D^{*+}\pi^-$ and $|a_1(K)|=\aOneKaon$ from $\Bb\to D^{*+}K^-$ are found based on BGL$(2,2,2)$ using Fermilab-MILC LQCD input. The first uncertainty is statistical from the hadronic branching fraction measurement, the second is systematic, and the third includes the semileptonic input uncertainty and all other Standard Model uncertainties. Compared to values found by BaBar data in Ref.~\cite{Fleischer:2010ca}, of $|a_1(\pi)|=0.98\pm0.04$ and $|a_1(K)|=0.96\pm0.05$, this corresponds to an improvement on the total uncertainty for the pion channel from $4.0\%$ to $2.2\%$ and for the kaon channel from $5.2\%$ to $2.7\%$, and a shift of the central values towards lower values.

For $|a_1(\pi)|$ and $|a_1(K)|$ the large observed deviation can imply a large, $13-16\%$ non-factorizable contribution to the matrix elements, new physics contributions to the Wilson Coefficients, \cite{Iguro:2020ndk,Cai:2021mlt} or both.
Theoretical analyses of non-factorizable contributions in $B^0 \to J/\psi K^{0}_{{\rm S}}$ decays suggest contributions of the size $O(10^{-3})$~\cite{Li_2007}, which is also in clear disagreement with the result obtained above.

The results for $|a_1(K)|^2/|a_1(\pi)|^2$ are listed in Table~\ref{tab:a1_over_a1_results}. The value of $|a_1(K)|^2/|a_1(\pi)|^2 = \aOneRatio$ is found to be consistent with $SU(3)$ symmetry. Furthermore, the ratio is calculated for different particle species, which also agree with $SU(3)$ symmetry. Systematic uncertainties related to $D^{*+}$ reconstruction cancel out as both measurements are performed with the same Belle data set.

\begin{table*}[htb]
	\centering
	\caption{The extracted values of $|a_1$($h_1$)$|^2/|a_1$($h_2$)$|^2$ for $h=\pi,~K, ~\rho,~K^*,~a_1$ using the three semileptonic input scenarios described in the text.}
	\begin{tabular}{ c | c | c   }
		Model & Ratio & Result \\
\hline
  BGL(2,2,2), F-MILC & $|a_1(K)|^2/|a_1(\pi)|^2$ & $1.066\pm0.054$ \\
  BGL(2,2,2), JLQCD & $|a_1(K)|^2/|a_1(\pi)|^2$ & $1.057\pm0.054$ \\
  CLN noHQS, JLQCD & $|a_1(K)|^2/|a_1(\pi)|^2$ & $1.061\pm0.054$ \\
\hline
  BGL(2,2,2), F-MILC & $|a_1(\rho)|^2/|a_1(\pi)|^2$ & $0.87\pm0.13$ \\
  BGL(2,2,2), JLQCD & $|a_1(\rho)|^2/|a_1(\pi)|^2$ & $0.86\pm0.13$ \\
  CLN noHQS, JLQCD & $|a_1(\rho)|^2/|a_1(\pi)|^2$ & $0.87\pm0.13$ \\
\hline
  BGL(2,2,2), F-MILC & $|a_1(K^{*})|^2/|a_1(\pi)|^2$ & $0.83\pm0.16$ \\
  BGL(2,2,2), JLQCD & $|a_1(K^{*})|^2/|a_1(\pi)|^2$ & $0.81\pm0.16$ \\
  CLN noHQS, JLQCD & $|a_1(K^{*})|^2/|a_1(\pi)|^2$ & $0.82\pm0.16$ \\
\hline
  BGL(2,2,2), F-MILC & $|a_1(a_1)|^2/|a_1(\pi)|^2$ & $1.23\pm0.28$ \\
  BGL(2,2,2), JLQCD & $|a_1(a_1)|^2/|a_1(\pi)|^2$ & $1.18\pm0.27$ \\
  CLN noHQS, JLQCD & $|a_1(a_1)|^2/|a_1(\pi)|^2$ & $1.20\pm0.27$ \\
\hline
  BGL(2,2,2), F-MILC & $|a_1(\rho)|^2/|a_1(K)|^2$ & $0.82\pm0.12$ \\
  BGL(2,2,2), JLQCD & $|a_1(\rho)|^2/|a_1(K)|^2$ & $0.81\pm0.12$ \\
  CLN noHQS, JLQCD & $|a_1(\rho)|^2/|a_1(K)|^2$ & $0.82\pm0.12$ \\
\hline
  BGL(2,2,2), F-MILC & $|a_1(K^{*})|^2/|a_1(K)|^2$ & $0.77\pm0.15$ \\
  BGL(2,2,2), JLQCD & $|a_1(K^{*})|^2/|a_1(K)|^2$ & $0.76\pm0.15$ \\
  CLN noHQS, JLQCD & $|a_1(K^{*})|^2/|a_1(K)|^2$ & $0.77\pm0.15$ \\
\hline
  BGL(2,2,2), F-MILC& $|a_1(a_1)|^2/|a_1(K)|^2$ & $1.15\pm0.26$ \\
  BGL(2,2,2), JLQCD& $|a_1(a_1)|^2/|a_1(K)|^2$ & $1.12\pm0.26$ \\
  CLN noHQS, JLQCD& $|a_1(a_1)|^2/|a_1(K)|^2$ & $1.13\pm0.26$ \\
\hline
  BGL(2,2,2), F-MILC& $|a_1(K^{*})|^2/|a_1(\rho)|^2$ & $0.94\pm0.23$ \\
  BGL(2,2,2), JLQCD& $|a_1(K^{*})|^2/|a_1(\rho)|^2$ & $0.94\pm0.23$ \\
  CLN noHQS, JLQCD& $|a_1(K^{*})|^2/|a_1(\rho)|^2$ & $0.94\pm0.23$ \\
\hline
  BGL(2,2,2), F-MILC& $|a_1(a_1)|^2/|a_1(\rho)|^2$ & $1.41\pm0.36$ \\
  BGL(2,2,2), JLQCD& $|a_1(a_1)|^2/|a_1(\rho)|^2$ & $1.38\pm0.37$ \\
  CLN noHQS, JLQCD& $|a_1(a_1)|^2/|a_1(\rho)|^2$ & $1.39\pm0.37$ \\
\hline
  BGL(2,2,2), F-MILC& $|a_1(a_1)|^2/|a_1(K^{*})|^2$ & $1.49\pm0.44$ \\
  BGL(2,2,2), JLQCD& $|a_1(a_1)|^2/|a_1(K^{*})|^2$ & $1.47\pm0.44$ \\
  CLN noHQS, JLQCD& $|a_1(a_1)|^2/|a_1(K^{*})|^2$ & $1.48\pm0.44$ \\
	\end{tabular}\label{tab:a1_over_a1_results}
\end{table*}

\section{Conclusion}
Measurements of branching fractions $\mathcal{B}(\pionmodemath) = \brpioncombabs$ and $\mathcal{B}(\kaonmodemath) = \brkaoncombabs$, as well as their ratio $R_{K/\pi} = \mathcal{B}(\kaonmodemath)/\mathcal{B}(\pionmodemath) = \ratiocombabs$, are presented. These are the first measurements of $\mathcal{B}(\pionmodemath)$ and $R_{K/\pi}$ from Belle and the most precise on $\mathcal{B}(\kaonmodemath)$, superseding previous Belle results. 
They are used to measure $|a_1(h)|$ with the aim of performing a precision test of QCD factorization.
The measurements of $|a_1(\pi)| =\aOnePion$ and $|a_1(K)| = \aOneKaon$ are the first performed within a single experiment, cancelling many experimental systematic uncertainties.
For the measurement of $|a_1(h)|$ we use BGL$(2,2,2)$ with Fermilab-MILC inputs as it provides the least model-dependent choice with the most robust error analysis.
All measured values of $|a_1(h)|$ are several standard deviations smaller than the theory prediction. 
In the ratios of $|a_1(h_1)|^2/|a_1(h_2)|^2$ for different particle types $h$, it is found that all of these are consistent with unity within one standard deviation. This indicates that $|a_1(h)|$ is indeed a universal quantity and $SU(3)$ symmetry holds in hadronic \B decays. 

\section{Acknowledgements}
This work, based on data collected using the Belle detector, which was
operated until June 2010, was supported by 
the Ministry of Education, Culture, Sports, Science, and
Technology (MEXT) of Japan, the Japan Society for the 
Promotion of Science (JSPS), and the Tau-Lepton Physics 
Research Center of Nagoya University; 
the Australian Research Council including grants
DP180102629, 
DP170102389, 
DP170102204, 
DE220100462, 
DP150103061, 
FT130100303; 
Austrian Federal Ministry of Education, Science and Research (FWF) and
FWF Austrian Science Fund No.~P~31361-N36;
the National Natural Science Foundation of China under Contracts
No.~11675166,  
No.~11705209;  
No.~11975076;  
No.~12135005;  
No.~12175041;  
No.~12161141008; 
Key Research Program of Frontier Sciences, Chinese Academy of Sciences (CAS), Grant No.~QYZDJ-SSW-SLH011; 
Project ZR2022JQ02 supported by Shandong Provincial Natural Science Foundation;
the Ministry of Education, Youth and Sports of the Czech
Republic under Contract No.~LTT17020;
the Czech Science Foundation Grant No. 22-18469S;
Horizon 2020 ERC Advanced Grant No.~884719 and ERC Starting Grant No.~947006 ``InterLeptons'' (European Union);
the Carl Zeiss Foundation, the Deutsche Forschungsgemeinschaft, the
Excellence Cluster Universe, and the VolkswagenStiftung;
the Department of Atomic Energy (Project Identification No. RTI 4002) and the Department of Science and Technology of India; 
the Istituto Nazionale di Fisica Nucleare of Italy; 
National Research Foundation (NRF) of Korea Grant
Nos.~2016R1\-D1A1B\-02012900, 2018R1\-A2B\-3003643,
2018R1\-A6A1A\-06024970, RS\-2022\-00197659,
2019R1\-I1A3A\-01058933, 2021R1\-A6A1A\-03043957,
2021R1\-F1A\-1060423, 2021R1\-F1A\-1064008, 2022R1\-A2C\-1003993;
Radiation Science Research Institute, Foreign Large-size Research Facility Application Supporting project, the Global Science Experimental Data Hub Center of the Korea Institute of Science and Technology Information and KREONET/GLORIAD;
the Polish Ministry of Science and Higher Education and 
the National Science Center;
the Ministry of Science and Higher Education of the Russian Federation, Agreement 14.W03.31.0026, 
and the HSE University Basic Research Program, Moscow; 
University of Tabuk research grants
S-1440-0321, S-0256-1438, and S-0280-1439 (Saudi Arabia);
the Slovenian Research Agency Grant Nos. J1-9124 and P1-0135;
Ikerbasque, Basque Foundation for Science, Spain;
the Swiss National Science Foundation; 
the Ministry of Education and the Ministry of Science and Technology of Taiwan;
and the United States Department of Energy and the National Science Foundation.
These acknowledgements are not to be interpreted as an endorsement of any
statement made by any of our institutes, funding agencies, governments, or
their representatives.
We thank the KEKB group for the excellent operation of the
accelerator; the KEK cryogenics group for the efficient
operation of the solenoid; and the KEK computer group and the Pacific Northwest National
Laboratory (PNNL) Environmental Molecular Sciences Laboratory (EMSL)
computing group for strong computing support; and the National
Institute of Informatics, and Science Information NETwork 6 (SINET6) for
valuable network support.

\bibliography{references}
\end{document}

%% file: pub611-orcid.tex
\noaffiliation
\author{J.-F.~Krohn\,\orcidlink{0000-0002-5001-0675}} 
\author{D.~Ferlewicz\,\orcidlink{0000-0002-4374-1234}} 
\author{P.~Urquijo\,\orcidlink{0000-0002-0887-7953}} 

  \author{I.~Adachi\,\orcidlink{0000-0003-2287-0173}} 
  \author{H.~Aihara\,\orcidlink{0000-0002-1907-5964}} 
  \author{S.~Al~Said\,\orcidlink{0000-0002-4895-3869}} 
  \author{D.~M.~Asner\,\orcidlink{0000-0002-1586-5790}} 
  \author{H.~Atmacan\,\orcidlink{0000-0003-2435-501X}} 
  \author{T.~Aushev\,\orcidlink{0000-0002-6347-7055}} 
  \author{R.~Ayad\,\orcidlink{0000-0003-3466-9290}} 
% ...
\collaboration{The Belle Collaboration}